\documentclass[preprint]{elsarticle}

\usepackage[english]{babel}
\usepackage{booktabs}
\usepackage{siunitx}
\usepackage[labelfont=bf]{caption}
\usepackage{tabularx}
\usepackage{mathtools, amsmath,amsthm,amsfonts,amssymb}
\usepackage[colorlinks=true, allcolors=blue]{hyperref}
\usepackage[strings]{underscore}
\usepackage{cases}
\usepackage[overload]{empheq}
\DeclarePairedDelimiter{\norm}{\lVert}{\rVert}
\DeclareMathOperator*{\argmin}{arg\,min}

\newcommand{\bs}[1]{\boldsymbol{#1}}
\newcommand{\restr}[1]{\rvert_{#1}}

\begin{document}

\begin{frontmatter}

\title{Exact Boundary Enforcement Along Implicit Geometries for Physics-Informed, Deep Learning Problems in Continuum Mechanics}

\author[1]{Cody Rucker\corref{c1}}
\cortext[c1]{Corresponding author}
\ead{crucker@uoregon.edu}

\author[1,2]{Brittany A. Erickson}

\affiliation[1]{
    organization={Department of Computer Science, University of Oregon},
    addressline={1477 E 13th Ave},
    postcode={97403},
    city={Eugene},
    country={United States of America}
    }

\affiliation[2]{
    organization={Department of Earth Sciences, University of Oregon},
    addressline={1272 University of Oregon},
    city={Eugene},
    postcode={97403},
    country={United States of America}
    }

\begin{abstract} 
Solutions to well-posed problems in continuum mechanics are continuously dependent upon prescribed boundary conditions. Because of this, variations in the enforcement of boundary data can impact the reliability of inversion techniques that rely on efficient and accurate forward models.
To this end, it is necessary to understand how specific boundary implementation techniques can affect the performance of a given forward model. 
Our work focuses on the impact that key modeling decisions have on physics-informed neural network (PINN) solutions for initial boundary value problems in continuum mechanics. 
By interpolating boundary data over implicit boundary representations, 
we measure the performance of a physics-informed neural network across different configurations of soft and hard boundary enforcement. 
We target the problem of elastodynamic plane-strain and present a method of hard-enforcement of traction conditions over arbitrary,
implicitly-defined, domain boundaries considering both first and second order formulations of the governing equations. 
We show that PINNs achieve a higher relative accuracy when solving the first-order plane strain problem 
and we observe a tradeoff between the final relative error and the total run time to complete training. 
This tradeoff is characterized by the number of hard and soft boundaries where, in the extremes, 
all soft-enforcement results in greater accuracy with a longer run time, 
while all hard-enforcement leads to lesser accuracy and a shorter run time. 
\end{abstract}

\begin{keyword}
physics-informed neural network \sep plane strain \sep exact boundary enforcement \sep forward problem \sep fully dynamic
\end{keyword}

\end{frontmatter}
\section{Context and Motivation}
When solving the partial differential equations (PDE) of continuum mechanics, sufficient care must be taken to ensure that the prescribed boundary conditions and constitutive laws yield a well-posed problem. 
Such well-posed problems are guaranteed to have unique solutions which depend continuously on their boundary conditions (BC) and thus proper boundary enforcement is necessary. 
This is particularly important in geophysical modeling where key system characteristics manifest as boundary conditions along subsurface structures. 
To characterize the Earth's subsurface and identify the drivers of geologic unrest then requires solutions to inverse problems that are built on efficient forward models that reliably and accurately enforce BCs. 
This drives modeling efforts which seek to understand, identify, and mitigate geologic hazards. 

The effectiveness of geologic hazard assessment and forecasting is critically dependent on the inference of subsurface characteristics drawn from measurements taken at the Earth's surface. 
Two of the most common observations that can be leveraged  in this regard come from geodesy, which documents surface deformation through time and seismology, which measures acoustic wave propagation within the Earth. 
These measurements form the backbone of both the  US Geological Survey's National Volcano Early Warning System\citep{ewert20182018} and the National Earthquake Hazard Reduction Programs\citep{hayes2024us} which both rely on modeling efforts to assimilate data into reliable solutions to inverse problems.
While there is still a need to expand monitoring infrastructure\citep{ANSS2017}, the next decade will likely see an unprecedented increase in indirect, surface observations and new tools may be necessary to effectively assimilate unconventional data sources into modeling efforts \citep{Bergen2019}.

Traditional numerical approaches for solving partial differential equations (PDE) 
(e.g. finite difference methods) have seen incredible growth in the past century, 
in particular in terms of convergence theory and high-performance computing. 
Traditional methods employ a variety of different techniques for boundary enforcement. For example
finite difference methods (FD) often impose Dirichlet boundary data via direct inclusion in a discrete solution array. 
Meanwhile, Neumann conditions can be enforced via one-sided difference schemes on the interior 
or by employing central difference schemes on ghost-nodes for improved accuracy over fewer nodes \citep{leveque2007finite}. 
The method of summation-by-parts simultaneous-approximation-term (SBP-SAT) 
imposes boundary conditions weakly through penalty methods in exchange for provable energy stability \citep{nordstrom2013summation, lundquist2014sbp}. 
In the finite element method (FEM), Neumann conditions are handled naturally 
and are automatically satisfied when formulating the variational form of the problem, 
while Dirichlet conditions are treated as constraints on the finite element space of approximate solutions. 
Here, accuracy of the approximate solution depends on the chosen finite element space and element size \citep{brenner2008mathematical}. 
Details surrounding boundary enforcement can significantly impact the properties of numerical approximations 
and is thus an important topic in the solution of initial-boundary value problems (IBVP). 
While much consideration has been given to this topic for traditional numerical approaches, 
the methods themselves are limited by their dependence on an underlying computational mesh. 
This mesh dependency introduces limitations when high resolution is needed, 
and while traditional methods are well-suited for solving forward problems, 
solving inverse problems require additional machinery and can be prohibitively expensive \citep{kern2016numerical, givoli2021tutorial}.  

Machine learning (ML), on the other hand, excels in the presence of large data 
and the approach of Physics-Informed Neural Networks (PINN)\citep{raissi2019physics} seamlessly integrates sparse and/or noisy data
while ensuring that model outcomes satisfy rigorous physical constraints to avoid observational bias 
that can lead purely data-driven models to predict physically unrealistic outcomes\citep{Zhao2023}. 
While they do not outperform traditional numerical methods (except in some high-dimensional settings where traditional methods become intractable) \citep{karniadakis2021physics},
PINNs may offer significant advantage over traditional numerical methods in two key ways. 
First, by re-framing the forward model as a parametric PDE, we can construct a reduced model 
whose evaluation at a given parameter is more efficient than computing a high-fidelity approximation to the forward problem \citep{dal2020data, bhattacharya2021model}. 
Second, both forward and inverse problems can be solved in the same computational framework 
without significant increase to computational load \citep{raissi2020hidden}. 
The latter of which implies that an efficient and accurate PINN that solves a forward problem 
can be immediately extended to address inverse problems 
where careful consideration of the forward solution benefits the inverse solution.

Though PINNs lack a robust error analysis that comes with traditional methods, error bounds have been described for some specific, limited cases \citep{kutyniok2022mathematics,de2021approximation,shin2020convergence, shin2020error,de2022error,mishra2022estimates_forward, mishra2022estimates_inverse} and they have seen applications in a variety of physical problems like Navier-Stokes \citep{wang2020multi, sun2020surrogate, jin2021nsfnets}, convection heat transfer \citep{cai2021physics}, solid mechanics, \citep{haghighat2021physics, goswami2020transfer} and the Euler equations \citep{mao2020physics} where they have shown promise. 
Efforts have been made to customize PINNs by specifying elements of the network training (e.g., activation functions, gradient optimization techniques, neural network architecture, and loss function structure \citep{cuomo2022scientific}) and by careful formulation of the loss function to reduce the order of the PDE resulting in a simpler learning problem \citep{ciarlet2002finite, ern2004theory}. 
The latter of these two approaches have resulted in deep learning analogues of Ritz \citep{yu2018deep} and Galerkin \citep{kharazmi2019variational, kharazmi2021hp, jagtap2020conservative} methods. 
Further hybridization with finite element methods (FEM) has led to the development of variationally correct loss functions, a class of trainable loss functions which provide upper and lower bounds on the approximation error (up to uniform constant factors)\citep{castillo2025dpg, bachmayr2025variationally}. 
Additionally, improved accuracy over other optimization algorithms can be achieved by incorporating a PDE energy into the backpropagation step \citep{muller2023achieving}. 
While this optimization scheme is intractable on a large network, \citet{mckay2025nearoptimal} developed a highly-accurate, memory-efficient implementation by sketching the natural gradient descent direction. 
 
Recently, efforts have been made to include exact matching of boundary conditions when solving problems in continuum mechanics.
\citet{sheikholeslami2025physics} tested soft and hard boundary enforcement on PINNs to solve the linear wave problem described by potential flow theory. 
\citet{Sahin2025} solved a 3D forward problem in contact mechanics on a simple domain using output transformations to enforce Dirichlet and Neumann boundary conditions as hard constraints. 
\citet{berrone2023enforcing} compared different approaches to enforcing Dirichlet conditions in PINNs and Variational PINNS (VPINNs) using soft-enforcement, hard-enforcement, and a variational imposition using Nitsche’s method. 
\citet{straub2025hard} used periodic Fourier feature embeddings to directly incorporate Neumann boundary conditions into the neural network’s architecture. 
This method relies on the fact that the normal derivative of the embedding vanishes near the prescribed boundary and is thus restricted to simple domains. 
For complex domain geometry, \citet{sobh2025pinn} hybridize FEM and PINNs using a domain decomposition. 
Here, FEM is used at Dirichlet boundaries and a PINN is used on the domain interior to ensure hard-enforcement of Dirichlet conditions in a variational formulation that naturally satisfies any Neumann conditions. 
In this work, we implement hard enforcement of boundary conditions by taking the approach detailed in \citet{sukumar2022exact} which interpolates boundary data using implicitly-defined approximate distance functions.

In general, PINNs are trained by minimizing a loss function that encodes governing equations, constitutive laws, and boundary conditions into a multi-objective loss function where each constraint is represented by its own loss component. 
This yields a complex training landscape where the magnitude of competing loss components may impact final accuracy of the trained network. 
Previous work has shown that hard-enforcement of temporal boundary conditions yielded a favorable performance increase when inferring subsurface friction parameters for an antiplane fault governed by a rate-and-state friction law\citep{rucker2024physics}.
We wish to extend the application and verification of PINN solutions to inverse problems by targeting laboratory fault experiments \citep[e.g.][]{cebry2024heterogeneous, Cebry2022, Mclaskey2017, McLaskey2019} 
which provide a small-scale, data-rich, environment for properly testing PINN inversion on real data sets. 
These laboratory experiments can be modeled by 2D dynamic plane strain which motivates the target problem of this work.
We focus on robust solutions to the forward problem, which is a key step prior to inversions. 
To better define key modeling decisions and their impact on PINN performance when solving IBVP, 
we investigate hard and soft enforcement of boundary conditions for various formulations of the 2D elastodynamic wave equation
governed by plane strain deformation (initial conditions are treated as boundary conditions at the temporal boundary $t=0$) 
and test PINN performance over various boundary types and enforcement configurations. 

Our goal is to explore how key modeling decisions may impact PINN training 
and highlight important considerations for constructing efficient deep-learning forward models. 
This paper is organized as follows: In Section~\ref{sec: PINN framework} we provide an overview of the physics-informed deep learning framework 
and PINN architecture for general initial–boundary-value problems 
in order to best describe the implementation to our application problem. 
In Section~\ref{sec: hard-enforcement} we provide the theory and construction of implicit boundary representations that yield exact boundary compliance via a generalized transfinite interpolation. 
In Section~\ref{sec: governing eqn} we formulate first and second-order plane strain problems and derive the generalized optimization problem for soft and hard boundary enforcement. 
In Section~\ref{sec: numerical tests} we report on PINN performance across various problem configurations verified against a manufactured solution. We conclude with a summary and discussion of future work in Section~\ref{sec: summary and future work}.

\section{Physics-Informed Deep Learning Framework}
\label{sec: PINN framework}
The physics governing motion in many applications in science and engineering give rise to PDE, for which, analytic solutions can be difficult to obtain (due, e.g. to complex material properties, boundary conditions, geometry). 
As a result, we commonly turn to numerical methods. 
One such method, popularized by \citet{raissi2019physics}, is the Physics-Informed Neural Network (PINN) which is a deep learning (DL) framework for approximating solutions to PDEs. 
Though this DL framework lacks the robust mathematical theory of the traditional methods, it shows particular promise in solving problems for which traditional numerical methods are ill-suited \citep{raissi2019physics}. 
The DL framework produces an analytic, closed-form solution, which is continuous and defined at every point in the domain (i.e. the solution is mesh-free). Moreover, forward and inverse problems can be solved in the same computational setting allowing for efficient forward models to be seamlessly incorporated into inversion efforts.

\subsection{Feed-forward Deep Neural Networks}
PINNs are extensions of a general feed-forward neural network. 
We let $\mathbf{x}\in \mathbb{R}^n$ and define a weighting matrix $W\in \mathbb{R}^{m\times n}$ and bias vector $b\in \mathbb{R}^m$. 
A single hidden layer of a neural network can be expressed as 
\begin{align}
    \ell(\mathbf{x}; \theta) = \varphi(W\mathbf{x} + b), \quad \text{where }\theta = (W, b),
\end{align}
and $\varphi$ is a known (nonlinear) activation function. 
Deep neural networks are obtained by repeated composition of hidden layers \citep{kollmannsberger2021deep}. 

We let positive integer $L$ be the deep neural network depth (i.e. number of hidden layers) and let $\{\varphi_i\}_{i=1}^L$ be a collection of activation functions along with a sequence of trainable network parameters $\{\theta_i\}_{i=0}^{L}$ where $\theta_k=(W_k, b_k)$ for each integer $0\leq k\leq L$. 
Here we assume each layer consists of $m$ neurons but this simplifying assumption may be omitted so long as one ensures that each weight and bias are of the correct dimension for matrix multiplication and addition, respectively. 
If $\mathbf{x}\in \mathbb{R}^n$ is the network input and network parameters $(W_0, b_0)\in \mathbb{R}^{m\times n}\times \mathbb{R}^m$, $(W_k, b_k)\in \mathbb{R}^{m\times m}\times \mathbb{R}^m$ for $1 \leq k<L$, and $(W_L, b_L)\in \mathbb{R}^{d\times m}\times \mathbb{R}^d$, then the recursive composition 
\begin{subequations}
    \begin{align}
	   \ell_0 &= \mathbf{x},\\
	   \ell_k &= \varphi_k (W_k \ell_{k-1} + b_k) \quad \text{ for } 0<k< L,
    \end{align}
    \label{Defn: recursive_NNet_def}%
\end{subequations}
defines a feed-forward, deep neural network $\mathcal{N}:\mathbb{R}^n \to \mathbb{R}^d$ (of depth $L$ and width $m$) with network parameters $\theta$ by $\mathcal{N}(\mathbf{x};\theta) = W_L\ell_{L-1} + b_L$. 

\subsection{PINN Architecture}
The PINN architecture is defined by extending the feed-forward deep neural network to enforce physical conditions set by an IBVP. 
For simplicity in the notation (as well as generality) we define this extension for a general IBVP and provide specific details for our application in Section~\ref{sec: governing eqn}. 
Consider the general operator form of an IBVP given by 
\begin{subequations}
    \begin{alignat}{2}
        \mathcal{L}\left[{\bf u}; \lambda\right] &= {\bf k}, \quad && \text{ in } \widehat{\Omega} , \label{eqn: ibvp(operator form)a}\\
        \mathcal{B}\left[{\bf u}; \lambda\right] &= {\bf g}, \quad &&\text{ on } \partial \widehat{\Omega},\label{eqn: ibvp(operator form)b}
    \end{alignat}
    \label{eqn: ibvp(operator form)}%
\end{subequations}
where $\widehat{\Omega}\subseteq \mathbb{R}^n$, with boundary (including internal interfaces) $\partial \widehat{\Omega}$. 
Vector ${\bf u}$ is the unknown solution and ${\bf g}$ is boundary data. 
The source term ${\bf k}$ encompasses external and internal body forces, and $\mathcal{L}$, $\mathcal{B}$ are differential and boundary operators parameterized by $\lambda$.

The PINN first approximates the solution to the IBVP \eqref{eqn: ibvp(operator form)} using a feed-forward, deep neural network. 
That is, we assume ${\bf u}({\bf x})\approx \mathcal{N}({\bf x}; \theta)$ and consider $N_\partial$ boundary points $\{{\bf x}^i_\partial, {\bf g}^i\}_{i=1}^{N_\partial}$ where ${\bf g}^i={\bf g}({\bf x}^i_\partial)$ for each $1\leq i \leq N_\partial$, and a set of $N_{{\widehat{\Omega}}}$ internal collocation points $\{ {\bf x}_{{\widehat{\Omega}}}^i, {\bf k}^i\}_{i=1}^{N_{\widehat{\Omega}}}$ with ${\bf k}^i={\bf k}({\bf x}^i_{\widehat{\Omega}})$ for $1\leq i \leq N_{\widehat{\Omega}}$. 
Throughout this work, collocation points are generated from uniform random sampling. 
From \eqref{eqn: ibvp(operator form)} we construct an objective function $MSE$ (based on the mean-square error) given by
\begin{align}
    MSE(\theta) = MSE_{{\widehat{\Omega}}}(\theta)+  MSE_\partial(\theta) ,
    \label{eqn: MSE}%
\end{align}
where
\begin{subequations}
    \begin{align}
        MSE_{{\widehat{\Omega}}}(\theta) &= \frac{1}{N_{{\widehat{\Omega}}}}\sum_{i=1}^{N_{{\widehat{\Omega}}}} \big |\mathcal{L}[\mathcal{N};\lambda]({\bf x}^i_{{\widehat{\Omega}}}; \theta) - {\bf k}^i \big |^2,\\[-5pt]
        MSE_\partial(\theta) &= \frac{1}{N_\partial}\sum_{i=1}^{N_\partial} \big |\mathcal{B}[\mathcal{N}; \lambda]({\bf x}^i_\partial; \theta) - {\bf g}^i \big |^2,\label{eqn: bdry loss}
    \end{align}
    \label{eqn: MSE loss}%
\end{subequations}
are the contributions to the total loss from the PDE \eqref{eqn: ibvp(operator form)a} and boundary conditions \eqref{eqn: ibvp(operator form)b}, respectively. 
Since our objective function \eqref{eqn: MSE} consists of multiple competing loss functions, it is referred to as a multi-objective loss. 
Solving \eqref{eqn: ibvp(operator form)} is done by minimizing \eqref{eqn: MSE} with respect to the network parameters $\theta$, typically done via an optimization algorithm that approximates the set of solution parameters $\theta^*$ such that $\theta^* = \argmin_\theta MSE(\theta)$. 
The inclusion of loss term~\eqref{eqn: bdry loss} imposes BC through \textit{soft enforcement}, i.e., the output of the neural network only approximately satisfies boundary conditions. 
This leads to imperfect satisfaction of the BC since minimization of \eqref{eqn: MSE} entails competition between enforcement of the BC and PDE. 
Alternatively, we may embed the boundary condition into the network's architecture to ensure exact enforcement of BC, a process known as \textit{hard enforcement}, which we now describe. 

\section{Hard-Enforcement of Boundary Conditions}
\label{sec: hard-enforcement}
\citet{lagaris1998artificial, lagaris2000neural} have presented a method of hard-enforcement of boundary conditions within the PINN framework by designing solution forms that automatically satisfy conditions at the domain boundary while training a neural network on the domain interior. 
As a straightforward example, suppose we wish to approximate some function $u$ subject to the conditions: $u(x,0) = u_0(x)$ and $u_t(x,0) = v_0(x)$. 
Then, given a neural network $\mathcal{N}(x,t;\theta)$, we may propose a trial solution of the form
\begin{align}
    U(x, t; \theta) = u_0(x) + tv_0(x) + t^2\mathcal{N}(x, t; \theta).\label{eqn: hard IC}
\end{align}
This ensures that $U(x, 0;\theta) = u_0(x)$ and $U_t(x, 0;\theta)=v_0(x)$ exactly, and the trainable term $\mathcal{N}$ in \eqref{eqn: hard IC} vanishes at the boundary $t=0$. 
This hard-enforcement approach can be generalized to enforce boundary conditions over regions of arbitrary topology via a generalized transfinite interpolation on implicitly-defined sets \citep{rvachev2001transfinite}. 
Our contributions are built on the work of \citet{sukumar2022exact} and the historical development of R-function theory \citep{shapiro1991theory, rvachev2001transfinite, shapiro2007semi}. 

\subsection{Transfinite Interpolation of Boundary Data using Inverse Distance Weighting}
Let $\Omega \subset \mathbb{R}^d$ be an open, bounded set with boundary $\partial \Omega$ partitioned by $\{\Gamma_i\}_{i=1}^n$. 
The distance from a given $\mathbf{x}\in \mathbb{R}^d$ and the set $S$ is defined by $d(\mathbf{x}, S)=\inf_{\mathbf{s}\in S}\norm{\mathbf{x}-\mathbf{s}}$. 
Subscripts are used to denote distance functions for a particular subset of the boundary, that is, $d_i(\mathbf{x})\coloneqq d(\mathbf{x}, \Gamma_i)$. 
For a set of prescribed values $f_i:\Gamma_i\to\mathbb{R}$ we construct an interpolating function as the linear combination of prescribed values $\{f_i\}_{i=1}^n$ with the basis $\{w_i\}_{i=1}^n$ formed by inverse distance weighting \citep{shepard1968two}. 
The result is the transfinite interpolant \citep{rvachev2001transfinite, sukumar2022exact}
\begin{align}
    f(\mathbf{x}) = \sum_{i=1}^n f_i w_i(\mathbf{x}), \quad w_i = \frac{d_i^{-\eta_i}}{\sum_{j=1}^n d_j^{-\eta_j}}=\frac{\prod_{j=1; j\neq i}^n d_j^{\eta_j}}{\sum_{k=1}^n \prod_{j=1; j\neq k}^n d_j^{\eta_j}},\label{eqn: interpolating function}
\end{align}
where each basis function $w_i$ is a positive, continuous function which forms a partition of unity $\left(\sum w_i(\mathbf{x})=1\right )$ and satisfies the interpolation condition $\left (w_i(\mathbf{x}_j)=\delta_{ij}\right )$. 
Exponents $\eta_i$ provide control over the differential properties of the interpolating function at boundary $\Gamma_i$. 
More specifically, for exponents $\eta_i>1$, the interpolant is differentiable (with vanishing normal derivatives) up to order $\eta_i-1$  at $\Gamma_i$  \citep{hoschek1993fundamentals}. 

From ~\eqref{eqn: interpolating function} we observe that distance functions can fully represent  geometric boundaries but some issues arise if we proceed by using exact distance functions. 
In general, distance fields must be computed numerically when working with boundaries defined by free-form curves or surfaces for which an iterative method is necessary due to geometric nonlinearity of the boundary \citep{upreti2014algebraic}. 
Moreover, exact distance functions may not be continuously differentiable. 
That is, any domain point that is equidistant to at least two points on the same boundary will produce a derivative discontinuity in the distance function. 
Because we aim to employ the transfinite interpolation \eqref{eqn: interpolating function} to solve IBVPs, we opt for approximate distance functions (ADF) which have a closed-form expression. 
While they behave like exact distances near the boundary, ADFs are constructed to be smooth, monotonically increasing functions of distance so that they still extend the influence of boundary terms on the domain interior in \eqref{eqn: interpolating function}.

\subsection{Approximate Distance Functions}
Approximation criteria are defined from characteristic properties of exact distance functions. 
Consider a boundary $\Gamma$ with unit inward normal $\bs{\nu}$ and distance function $d_\Gamma(\mathbf{x}) =d(\mathbf{x}, \Gamma)$. 
Observe that distance functions implicitly define their associated boundaries by their zero curve. 
That is, $d_\Gamma(\mathbf{x})=0$ $\forall \mathbf{x}\in\Gamma$ is a characteristic property of distance functions that ADFs should replicate. 
Additionally, for any $\mathbf{x}\in \Omega$, the nearest point $\mathbf{y}\in\Gamma$ is given by the orthogonal projection of $\mathbf{x}$ on to $\Gamma$ so that $d_\Gamma(x) = d(x,y)=\norm{x-y}$. 
Note that this means $\mathbf{x}-\mathbf{y}$ is orthogonal to $\Gamma$ at $\mathbf{y},$ where the unit normal $\bs{\nu}_{\mathbf{y}}$ gives the direction of maximal increase at $\mathbf{x}$ and thus $\bs{\nu}_\mathbf{y}=\alpha(\mathbf{x}-\mathbf{y})$ for some $\alpha>0$. 
This lets us reduce distance perturbations in the direction normal to $\Gamma$ at $\mathbf{y}$ to $d_\Gamma(\mathbf{x} + \bs{\nu}_\mathbf{y})=(1+\alpha)\norm{\mathbf{x}-\mathbf{y}}$. 
Thus, the directional derivative 
\begin{align}
    D_{\bs{\nu}_\mathbf{y}}d_\Gamma(\mathbf{x}) = \lim_{h\to0}\dfrac{d_\Gamma(\mathbf{x} + h\bs{\nu}_\mathbf{y}) - d_\Gamma(\mathbf{x})}{h\norm{\bs{\nu}_\mathbf{y}}}=1.
\end{align}
Finally, since $D_{\bs{\nu}_\mathbf{y}}d_\Gamma = \nabla d_\Gamma \cdot \bs{\nu}_\mathbf{y}$ we can conclude that $\norm{\nabla d_\Gamma}=1$. 
This property implies that at the boundary $\Gamma$, distance function $d_\Gamma$ is linear in the direction of the unit normal $\bs{\nu}$ and thus normal derivatives beyond order one vanish. 
With these properties laid out, the function $\omega$ is said to be \textit{normalized up to $m^\text{th}$ order} when the following conditions are satisfied:
\begin{align}
    \omega_{ |\Gamma}=0, \quad 
    \dfrac{\partial \omega}{\partial \nu}_{ |\Gamma} = 1, \quad \dfrac{\partial^k \omega}{\partial \nu^k}_{ |\Gamma} = 0,\quad k=2, 3,\dots,m ,\label{eqn: order m normalization}
\end{align}
where $\bs{\nu}$ is the inward normal of the boundary $\Gamma$ which is defined implicitly by $\omega=0$. 
In this way, $\omega$ acts as the $m^\text{th}$ order approximation of the distance function near boundary $\Gamma. 
$ We now give a method of constructing ADFs for straight-line segments that are normalized to first order. 
For brevity, we omit the details behind R-function theory that allow for this construction but provide a brief introduction in \ref{sec: R-function theory} and refer to \citet{rvachev2001transfinite} and \citet{sukumar2022exact} for details. 
In the following section we use two key ideas from R-function theory. First, a single implicit function can be computed to represent objects constructed from Boolean set operations on halfspace primitives. Second, many R-functions preserve the order of normalization of their arguments. 
With this, we show that an unbounded implicit line (normalized to order 1) can be trimmed using a a convex trimming region (also normalized to order 1) to produce an ADF for the segment of the line bounded by the trimming region that is normalized to order 1\citep{shapiro1991theory, rvachev2001transfinite, shapiro1999implicit}.

\subsection{Approximate distance to a trimmed line segment}
The line segment joining points $\mathbf{x}_1 = (x_1, y_1)$ and $\mathbf{x}_2 =(x_2, y_2)$ has center point $\mathbf{x}_\text{c}=(\mathbf{x}_1 + \mathbf{x}_2)/2$ and length $L= \norm{\mathbf{x}_2 - \mathbf{x_1}}$ admits the signed distance function from $\mathbf{x}\in \mathbb{R}^2$ to the line passing through $\mathbf{x}_1$ and $\mathbf{x}_2$ given by
\begin{align}
    f\coloneqq \frac{(x-x_1)(y_2-y_1) - (y-y_1)(x_2-x_1)}{L}. \label{eqn: implicit line}
\end{align}
The ADF for the line segment is given by the intersection of the infinite line with a circle of radius $L/2$ centered at $\mathbf{x}_c$. 
We thus define a convex trimming region by $t\geq 0$ for the trimming function
\begin{align}
    t\coloneqq \frac{1}{L}\left[ \left( \frac{L}{2}\right )^2 - \norm{\mathbf{x} - \mathbf{x}_\text{c}} \right],\label{eqn: circular trim}
\end{align}
which is normalized to first order \citep{rvachev2001transfinite}. 
The ADF for the line segment is then given by the normalized trimming procedure described in \citet{shapiro1999implicit}
\begin{align}
    \phi \coloneqq \sqrt{f^2 + \left ( \frac{\varphi - t}{2}\right )^2}, \quad \varphi = \sqrt{t^2 + f^4}. 
 \label{eqn: normalized trim}
\end{align}

\begin{figure}
    \centering
    \includegraphics[width=1.0\linewidth]{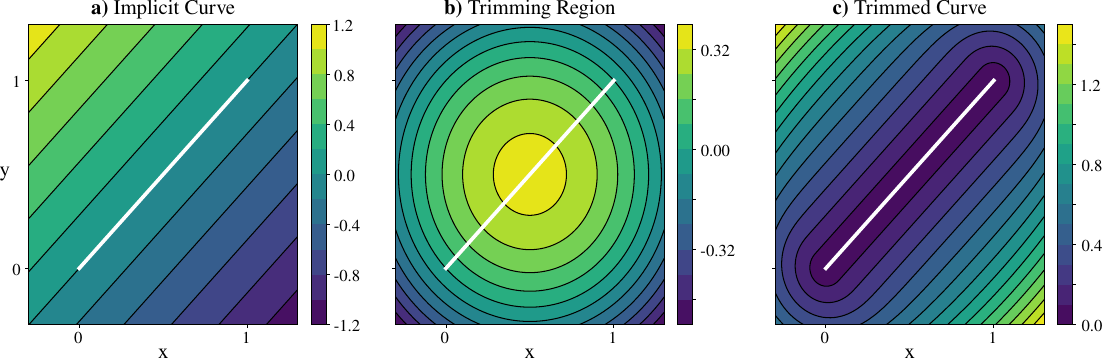}
    \caption{Construction of an approximate distance function for the line segment joining points $A=(0,0)$ and $B=(1,1)$. a) depicts the signed distance function \eqref{eqn: implicit line} to the line $\overline{AB}$; b) depicts the trimming function \eqref{eqn: circular trim}; and c) depicts the resulting approximate distance function \eqref{eqn: normalized trim} to line segment $\overline{AB}$.}
    \label{fig: adf construction}
\end{figure}

Figure~\ref{fig: adf construction} provides an illustration of $f, t, $ and $\phi$ for constructing an ADF for the line segment joining points $(0,0)$ and $(1,1)$.
Trimmed entity $\phi$ inherits differential properties from the functions $f,t$ and trimming procedure \eqref{eqn: normalized trim}\citep{shapiro1999implicit}. 
This can be seen by considering $\varphi=\vert t \vert$ in \eqref{eqn: normalized trim} which has a derivative discontinuity at the boundary of the trim region $t=0$. 
The construction laid out above yields an ADF which is $C^2$ at every point outside of its zero set and is fully normalized at the regular points of the line segment \citep{biswas2004approximate, rvachev2001transfinite, sukumar2022exact}.

This demonstrates the construction of ADFs for linear boundary segments which is sufficient for representing the domain boundaries present in this work. 
However, ADFs can be constructed for more general curves by applying \eqref{eqn: normalized trim} to appropriately-defined implicit and trimming functions. 
Details for constructing ADFs for Bezier curves are provided in \ref{sec: A bezier}, and we refer the reader to \citet{upreti2014algebraic} for further details.
For the remainder of this work, when we refer to the transfinite interpolation \eqref{eqn: interpolating function} it is assumed that ADFs are used in place of exact distance functions. This completes our discussion on boundary construction since the interpolation bases $w_i$ in \eqref{eqn: interpolating function} are fully specified by inverse distance weighting on a set of boundary ADFs. 
Because we aim to leverage \eqref{eqn: interpolating function} for exact enforcement of traction conditions, we must consider how to prescribe $f_i$ at boundary $\Gamma_i$ to interpolate both function values and derivatives.

\subsection{Transfinite interpolation of normal derivatives}
Taylor series are a natural choice for interpolating a function's derivatives at a point so we specify boundary data functions to be Taylor series approximations in some neighborhood of the boundary. For clarity, we first demonstrate the method for interpolating derivative values in one dimension. Consider a function $u:\mathbb{R}\to\mathbb{R}$ which satisfies 
\begin{align}
    u(a) = f, \quad u'(a)=f', 
\end{align}
then $u$ can be represented in a neighborhood of the point $a\in\mathbb{R}$ by a Taylor series expansion
\begin{align}
    u(x) = f + f'(x-a) + \mathcal{O}\left ((x-a)^2 \right ).
\end{align}
Prescribing $f_a(x)=f+f'(x-a)$ in a transfinite interpolation scheme then interpolates both value and derivative data. 
This approach can be applied to problems in higher dimensions by specifying a directional Taylor series expansion with respect to a sufficiently normalized boundary ADF. 
We provide the generalized Taylor series expansion for representing functions in the neighborhood of a boundary described by the zero set of a normalized ADF.

Given a bounded domain $\Omega \subset \mathbb{R}^3$ with boundary segment $\Gamma$ represented by ADF $\gamma$ that is normalized up to order $m$, if function $\mathbf{u}:\Omega\to\mathbb{R}^2$ satisfies the conditions
\begin{align}
    \mathbf{u}(\mathbf{x})_{ |\Gamma} = f_0(\mathbf{x}),\quad \dfrac{\partial^k\mathbf{u}}{\partial\bs{\nu}^k}_{ |\Gamma}=f_k(\mathbf{x}), \quad (k=1, 2, \dots,m), 
\end{align} 
then $\mathbf{u}$ can be represented in a neighborhood of the boundary $\Gamma$ by 
\begin{align}
    \mathbf{u} = f_0^* + \sum_{k=1}^m \frac{1}{k!}f_k^* \gamma^k + \mathcal{O}(\gamma^{m+1})\label{eqn: gen taylor}
\end{align}
where for each $k=0,1,\dots,m$, $f^*_k,$ are \textit{normalizers} of $f_k$ with respect to $\gamma$. Here $f_k$ represents prescribed values and derivatives of order $k$ on $\Gamma$. Normalizers should agree with prescribed values at the boundary (i.e., $f^*_k(\mathbf{x})=f_k(\mathbf{x})$ for $\mathbf{x}\in \Gamma$) but, due to their role as coefficients in the Taylor expansion~\eqref{eqn: gen taylor}, they should be constant along the direction normal to the boundary. This last part ensures that the partial sum in \eqref{eqn: gen taylor} accurately represents $\mathbf{u}$ in a neighborhood of the boundary.

A normalizer $f^*$ is constructed by linearizing $f$ in the neighborhood $\gamma = 0$ and subtracting the variation in the normal direction $\boldsymbol{\nu}=\nabla \gamma$ to get
\begin{equation}\label{eqn: general normalizers}
    \begin{aligned}
        f^* &= f - \gamma \big [\nabla \gamma \cdot \mathbf{J}_f \big ] + \mathcal{O}(\gamma^2) \\
            &= f + \gamma\mathbf{D}_f^{\gamma} +  \mathcal{O}(\gamma^2),
    \end{aligned}
\end{equation}
where $\mathbf{J}_f$ denotes the Jacobian of $f$, outward unit normal $\mathbf{n}=-\nabla \gamma$, and 
$\mathbf{D}_f^{\gamma} = - \mathbf{J}_f \cdot \nabla \gamma$ defines a differential operator that acts in the outward normal direction to the boundary defined by $\gamma=0$. 

To illustrate how \eqref{eqn: gen taylor} and \eqref{eqn: general normalizers} are used to define boundary data functions, we consider the following specifications for $\mathbf{u}$ on boundary $\Gamma$:  
\begin{align}
    \mathbf{u}\restr{\Gamma} = \mathbf{g}, \quad \left. \dfrac{\partial \mathbf{u}}{\partial \boldsymbol{\nu}}\right\restr{\Gamma}  =\mathbf{h}.\label{eqn: specified data}
\end{align}
Then \eqref{eqn: gen taylor} gives us
\begin{align}
    \mathbf{u} = \mathbf{g}^* + \gamma \mathbf{h}^* + \mathcal{O}(\gamma ^2),
\end{align}
where normalizers $\mathbf{g}^*, \mathbf{h}^*$ are given by \eqref{eqn: general normalizers} to be 
\begin{subequations}
    \begin{align}
        \mathbf{g}^* &= \mathbf{g} + \gamma \mathbf{D}_\mathbf{g}^{\gamma} + \mathcal{O}(\gamma^2),\\
        \mathbf{h}^* &= \mathbf{h} + \gamma \mathbf{D}_\mathbf{h}^{\gamma} + \mathcal{O}(\gamma^2).\label{eqn: normal bc}
    \end{align}
\end{subequations}
Thus, we get
\begin{equation}
    \begin{aligned}
        \mathbf{u} &= \mathbf{g} + \gamma \mathbf{D}_\mathbf{g}^{\gamma} + \gamma \mathbf{h} + \gamma^2 \mathbf{D}_\mathbf{h}^{\gamma} + \mathcal{O}(\gamma^2)\\
        &= \mathbf{g} + \gamma (\mathbf{D}_g^{\gamma}  + \mathbf{h}) + \mathcal{O}(\gamma^2).
    \end{aligned}
\end{equation}
Note that this enforces the directional derivative condition in \eqref{eqn: specified data} with respect to the inward unit normal $\boldsymbol{\nu}$. 
 Neumann conditions are typically specified with respect to the outward unit normal $\mathbf{n}$, thus if we want to enforce the Neumann condition $\partial \mathbf{u}/\partial \mathbf{n}  = \hat{\mathbf{h}}$ we note that $\partial \mathbf{u} / \partial\boldsymbol{\nu} = \mathbf{h} = -\hat{\mathbf{h}}$ so that
\begin{align}
    \mathbf{u} = \mathbf{g} + \gamma \big [\mathbf{D}_g^{\gamma} - \hat{\mathbf{h}} \big ] + \mathcal{O}(\gamma^2)\label{eqn: normal soln form}
\end{align}
is a general normalizing structure for interpolating both function values and normal derivatives at boundary $\Gamma$. 
Equation~\eqref{eqn: normal soln form} can be employed to enforce Dirichlet or Neumann conditions when approximating solution $\mathbf{u}$. 

\subsubsection{Dirichlet Conditions}
When specifying boundary data functions that interpolate prescribed values at boundary $\Gamma$, we use a boundary subscript to denote a truncated Taylor series. 
This helps distinguish boundary data functions from the target function. 
As an example of this, for the case of Dirichlet conditions $\mathbf{u}\restr{\Gamma}=\mathbf{g}$, the general normalizer~\eqref{eqn: normal soln form} yields the Dirichlet-specific boundary data function  
\begin{align}
    \mathbf{u}_\Gamma = \mathbf{g}.\label{eqn: specific dirichlet normalizer}
\end{align}

\subsection{Neumann Conditions}
For data prescribed on normal derivatives of the function (with respect to the outward unit normal $\mathbf{n})$, we still specify function values at $\Gamma$ but do so by prescribing values from a numerical approximation. In our case, a trainable network $\mathcal{N}$ is employed to approximate solutions on the domain interior so that $\mathcal{N}\approx\mathbf{u}\restr{\Omega \setminus \partial \Omega}$. 
Then function values can be prescribed by extending the network approximation to the boundary enabling exact enforcement of Neumann conditions. 
That is, by  prescribing $\mathbf{u}\restr{\Gamma}=\mathcal{N}$, and $(\partial \mathbf{u}/\partial \mathbf{n})\restr{\Gamma}=\hat{\mathbf{h}}$, the general form for normalizers~\eqref{eqn: normal soln form} yields boundary data 
 \begin{align}
     \mathbf{u}_\Gamma = \mathcal{N} + \gamma \big[\mathbf{D}^\gamma_{\mathcal{N}} - \hat{\mathbf{h}} \big] \label{eqn: specific neumann normalizer}
 \end{align}
for enforcing Nuemann conditions. With this the transfinite interpolation \eqref{eqn: interpolating function} can be supplied with appropriate ADFs and boundary data functions to ensure exact enforcement of initial and boundary data. We now append the boundary interpolation function with a trainable network on the domain's interior to form a generalized solution structure for IBVP.

\subsection{Generalized Solution Form of IBVP}
The generalized solution form for an IBVP using the R-function method, combines transfinite interpolation \eqref{eqn: interpolating function} (to interpolate boundary data) with an interior approximation term that vanishes on the boundary \citep{rvachev1995r, rvachev2000completeness}. 
Consider a boundary partition $\{\Gamma_i\}_{i=1}^m$ with associated ADF representations $\{\gamma_i\}_{i=1}^m$ on which some function $\mathbf{u}$ is prescribed the conditions:
\begin{align}
    \mathbf{u}\restr{\Gamma_i}=\mathbf{g}_i, \, (i=1,\dots,k), \quad \left. \dfrac{\partial\mathbf{u}}{\partial\mathbf{n}}\right \rvert_{\Gamma_j}=\mathbf{h}_j, \,(j=k+1,\dots,m).\label{eqn: general BCs }
\end{align}
Then, for an approximating neural network $\mathcal{N}$, an approximate solution form satisfying conditions \eqref{eqn: general BCs } is given by 
\begin{align}
    \mathbf{u} \approx \sum_{i=1}^mu_iw_i + \mathcal{N}\prod_{i=1}^m\gamma_i^{\eta_i} \label{eqn: general solution form}
\end{align}
where $w_i, \eta_i$ are the inverse distance weights and interpolant control exponents given in \eqref{eqn: interpolating function} where ADF $\gamma_i$ are used in place of exact distances and normalizers $u_i$ are determined from equations~\eqref{eqn: specific dirichlet normalizer} and \eqref{eqn: specific neumann normalizer} to be
\begin{align}
    u_i = 
    \begin{cases}
        \mathbf{g}_i, \quad& i=1,\dots,k\\
        \mathcal{N} + \gamma_i(\mathbf{D}_\mathcal{N}^{\gamma_i}-\mathbf{h}_i), \quad &i=k+1,\dots,m \label{eqn: general solution normalizers}
    \end{cases}.
\end{align}
Thus equations~\eqref{eqn: general solution form} and \eqref{eqn: general solution normalizers} provide a general solution form for approximating a function constrained by \eqref{eqn: general BCs }. 
We will now employ this solution form in a physics-informed learning problem that is generalized over hard and soft boundary enforcements. 
Furthermore, the normalizer for interpolating first-order derivatives in \eqref{eqn: general solution normalizers} requires that the condition be specified on the Jacobian of the network. 
In Section~\ref{sec: governing eqn} we show that hard-enforcement can be extended to more complex first-order conditions (e.g., traction conditions) by deriving traction-compliant Jacobian terms from the constitutive law and traction boundary condition. 
With the hard-enforcement of BC provided by \eqref{eqn: general solution form}, we can now proceed by defining the elastodynamic wave equation governed by plane strain motion which will be solved via a generalized solution form.

\section{Governing Equations}\label{sec: governing eqn}
We investigate the impact that specific modeling choices have on the performance of a trainable neural network while learning the solution to an initial boundary value problem. 
Motivated by sliding-block laboratory experiments \citep{cebry2024heterogeneous, Cebry2022, Mclaskey2017, McLaskey2019}, we target the problem of fully dynamic, 2D plane strain deformation (in both first and second order forms) and testing model performance across two types of IBVP while varying boundary enforcement type. 
Here we first present the IBVPs that will be used in Section~\ref{sec: numerical tests} to probe PINN performance with different types of boundary condition imposition. 
 
For notational convenience we denote kinematic and constitutive relations as operators. 
That is, for $\Omega\subset\mathbb{R}^2$ a bounded domain with boundary $\partial\Omega$ along with a time interval $t\geq0$,
a sufficiently smooth function $\bs{\psi}:\Omega\times[0,t]\to \mathbb{R}^3$ can be associated with stress and strain tensors, $\bs{\sigma}^{\bs{\psi}}, \bs{\varepsilon}^{\bs{\psi}}$, by
\begin{subequations}
    \begin{align}
        \sigma_{ij}^{\bs{\psi}} &= 2\mu \varepsilon_{ij}^{\bs{\psi}} + \lambda \varepsilon_{kk}^{\bs{\psi}} \delta_{ij},\label{eqn: order2 hooke}\\
        \varepsilon_{ij}^{\bs{\psi}} &= \frac{1}{2} \left (\psi_{i, j} + \psi_{j, i}\right ),\label{eqn: kinematic equations}
    \end{align}
\end{subequations}
where summation is implied by repeated indices, $\delta$ is the Kronecker delta, $\mu$ shear modulus, and $\lambda$ Lam\'e's first parameter. 
Here we use the convention that a comma in the subscript denotes partial differentiation with respect to the trailing indices. For displacement $\mathbf{u}=(u_x, u_y, u_z)$, the plane strain assumption supposes that displacements are restricted to the $x,y$ plane so that $u_z=0$ and there are no variations in the $z$-direction (i.e., $u_{x,z}=u_{y,z}=0$). This results in $\varepsilon^\mathbf{u}_{xz}=\varepsilon^\mathbf{u}_{xy}=\varepsilon^\mathbf{u}_{zz}=0$. Thus, we need only consider two components of displacement $\mathbf{u}=(u_x, u_y)$ with three components of strain $\bs{\varepsilon}^\mathbf{u}=(\varepsilon^\mathbf{u}_{xx}, \varepsilon^\mathbf{u}_{yy}, \varepsilon^\mathbf{u}_{xy})$ and stress  $\bs{\sigma}^\mathbf{u}=(\sigma^\mathbf{u}_{xx}, \sigma^\mathbf{u}_{yy}, \sigma^\mathbf{u}_{xy})$. With this assumption we proceed with the statement of the second-order wave equation. 

\subsection{Second-Order Form} 
For displacement vector $\mathbf{u}:\Omega\times[0,t] \to \mathbb{R}^2$ the 2D planestrain equations governing motion in the domain are given by
\begin{align}
    \rho \mathbf{u}_{tt} - \nabla \cdot{\bs{\sigma}}^\mathbf{u} = \mathbf{F}\label{eqn: 2nd-order governing eq}
\end{align}
where $\rho$ is a material density and $\mathbf{F}$ a numerical source for verifying accuracy against a manufactured solution (see Section~\ref{sec: 2nd order numerics}). 
Ensuring well-posedness of the problem requires that \eqref{eqn: 2nd-order governing eq} be supplied with sufficient boundary conditions. 
In this work we will consider both Dirichlet and Neumann boundaries so we let $\Gamma_D\cup\Gamma_N = \partial\Omega$ each be the portion of the boundary associated with Dirichlet and Neumann conditions, respectively with $\Gamma_D\cap\Gamma_N=\emptyset$. We supplement \eqref{eqn: 2nd-order governing eq} with boundary conditions
\begin{subequations}
    \begin{alignat}{3}
        \mathbf{u} &= \mathbf{g}(\mathbf{x}, t) \quad &&\mathbf{x}\in \Gamma_D, &&\quad t>0,\\
        \bs{\sigma}^{\mathbf{u}}\cdot \mathbf{n} &= \mathbf{h}(\mathbf{x}, t) \quad &&\mathbf{x}\in \Gamma_N, &&\quad t>0, \label{eqn: order2 traction BC}\\
        \mathbf{u} &= \mathbf{u}_0(\mathbf{x}) \quad &&\mathbf{x}\in \Omega, &&\quad t=0,\\
        \mathbf{u}_t &= \mathbf{v}_0(\mathbf{x}) \quad &&\mathbf{x}\in \Omega, &&\quad t=0,
    \end{alignat}\label{eqn: 2nd-order bc}%
\end{subequations}
to ensure the problem is well-posed.
When specifying a loss function to be minimized during training we must construct a loss component from \eqref{eqn: 2nd-order governing eq} that is used to train the network to satisfy the PDE on the domain interior. 
Additionally, each of the boundary conditions in \eqref{eqn: 2nd-order bc} can either be enforced via additional loss terms (soft-enforcement) or by constructing a trial function to exactly satisfy data at the boundary (hard-enforcement). 
Let $\{\Gamma_i\}_{i=1}^K$ be the collection of boundary segments where hard-enforcement is implemented and let $\{\overline{\Gamma}_i\}_{i=1}^M$ denote the remaining boundary segments where soft-enforcement is used. 
For boundary $\Gamma_i$ we let $\phi_i $ be an associated approximate distance normalized up to order $1$, and take neural network $\mathcal{N}$ so that the trial function  $\Phi\approx \mathbf{u}$ is defined by
\begin{align}
    \Phi = \sum_{i=1}^Kf_iw_i + \mathcal{N} \prod_{i=1}^K\phi_i^{\eta_i},\label{eqn: 2nd order trial function}%
\end{align}%
where $w_i$ are the inverse distance weights given by \eqref{eqn: interpolating function} using approximate distances $\phi_i$ instead of exact distances and exponents $\eta_i$, described in Section~\ref{sec: hard-enforcement}, control the differentiability of the interpolant at boundary $\Gamma_i$. 
The normalizers $f_i$ vary depending on the the boundary type at boundary $\Gamma_i$ which, in general, are given by
\begin{subequations}
    \begin{empheq}[left={f_i=\empheqlbrace}]{alignat=2}
        &\mathbf{g} \quad &&\text{ if } \Gamma_i\subseteq \Gamma_D\\
        &\mathcal{N} + \phi_i \left (\left (\mathbf{J}_\mathcal{N} - \mathbf{H}\right )\cdot \mathbf{n} \right) \quad &&\text{ if } \Gamma_i \subseteq \Gamma_N,\label{eqn: jacobian penalty}\\
        &\mathbf{u}_0 + \phi_i\mathbf{v_0}  &&\text{ if } \Gamma_i\subseteq \Omega \times \{0\},
    \end{empheq}
\end{subequations}
where the network Jacobian $\mathbf{J}_\mathcal{N}$ is penalized by $\mathbf{H}$ in \eqref{eqn: jacobian penalty} to implicitly satisfy the traction BC~\eqref{eqn: order2 traction BC}. This is done by first computing a traction-compliant stress $\mathbf{s}$ from a single component of the network stress using \eqref{eqn: order2 traction BC}. This gives 
\begin{subequations}
    \begin{align}
        s_{yy} &= 2\mu\mathcal{N}_{y,y} + \lambda \left (\mathcal{N}_{x,x} + \mathcal{N}_{y,y} \right ),\label{eqn: traction-compliant a}\\
        s_{xy} &= (h_y - s_{yy}n_2) / n_1,\label{eqn: traction-compliant b}\\
        s_{xx} &= (h_x - s_{xy}n_2) / n_1,\label{eqn: traction-compliant c}
    \end{align}\label{eqn: traction-compliant stress}%
\end{subequations}
so that $\mathbf{s}\cdot\mathbf{n}=\mathbf{h}$.
Note that network strains strains may also be specified with respect to the Jacobian as $\bs{\varepsilon}^{\mathcal{N}} = \left (\mathbf{J}_{\mathcal{N}} + \mathbf{J}_{\mathcal{N}}^T \right )/2$, which lets us specify a traction-compliant strain $\bs{\epsilon}=(\mathbf{H} + \mathbf{H}^T)/2$ which must satisfy
\begin{align}
    s_{ij} = 2\mu \epsilon_{ij} + \lambda \epsilon_{kk}\delta_{ij}.\label{eqn: traction-compliant strain}
\end{align}
To clarify, \eqref{eqn: traction-compliant strain} specifies that strains computed from $\mathbf{H}$ are related to the traction-compliant stress~$\mathbf{s}$ from \eqref{eqn: traction-compliant stress} via Hooke's law. 
So far, we have employed one equation from Hooke's Law in \eqref{eqn: traction-compliant a} with two equations from the traction conditions \eqref{eqn: traction-compliant b}, \eqref{eqn: traction-compliant c}.
Using the other two equations gained from Hooke's law 
\begin{align}
    \sigma^\mathcal{N}_{xx} = 2\mu \mathcal{N}_{x, x} + \lambda (\mathcal{N}_{x, x} + \mathcal{N}_{y, y}),  \quad \sigma^\mathcal{N}_{xy} = \mu (\mathcal{N}_{x, y} + \mathcal{N}_{y, x}),
\end{align}
we can solve for the first column of $\mathbf{J}_\mathcal{N}$ by 
\begin{align}
    \mathcal{N}_{x,x} = (\sigma_{xx}^\mathcal{N} - \lambda \mathcal{N}_{y,y})/(2\mu + \lambda), \quad 
    \mathcal{N}_{x,y} = (\sigma_{xy}^\mathcal{N}/\mu - \lambda \mathcal{N}_{x,y}).\label{eqn: network jacobian col}
\end{align}
Substituting traction-compliant stress $\mathbf{s}$ from \eqref{eqn: traction-compliant stress} in to  \eqref{eqn: network jacobian col} then yields the appropriate penalty 
\begin{align}
    \mathbf{H} &= \begin{bmatrix}
        (s_{xx} - \lambda \mathcal{N}_{y,y} )/(2\mu + \lambda) & \mathcal{N}_{x,y}\\
        (s_{xy}/\mu) - \mathcal{N}_{x,y} & \mathcal{N}_{y,y}
    \end{bmatrix}.
\end{align}
$\mathbf{H}$ can be verified by taking $\bs{\epsilon}=(\mathbf{H} + \mathbf{H}^T)/2$ and checking traction-compliance with \eqref{eqn: traction-compliant strain}. 
Thus, the first two columns of $\mathbf{J}_\mathcal{N}$ can be penalized to ensure traction-compliance at Neumann boundaries.

With this, we can now state the optimization problem to be solved by training network $\mathcal{N}$. 
Consider interior data given by the collocation points $\{\mathbf{z}_i\}_{i=1}^N\subset \Omega \times (0,t]$, Dirichlet boundary points $\{\hat{\mathbf{z}}_i\}_{i=1}^{M_D}\subset (\overline{\Gamma}\cap\Gamma_D)\times(0,t]$, Neumann boundary points $\{\bar{\mathbf{z}}_i\}_{i=1}^{M_N}\subset (\overline{\Gamma}\cap\Gamma_N)\times(0,t]$, and temporal boundary points $\{\tilde{\mathbf{z}}_i\}_{i=1}^{L}\subset \Omega \times \{0\}$. 
Then the component mean squared error (MSE) loss functions corresponding to governing equations~\eqref{eqn: 2nd-order governing eq} and boundary conditions \eqref{eqn: 2nd-order bc} for approximate trial solution \eqref{eqn: 2nd order trial function} are
\begin{subequations}
    \begin{align}
        MSE_{PDE} &= \frac{1}{N}\sum_{i=1}^N\norm{\rho \Phi_{tt}(\mathbf{z}_i) - \nabla \cdot \bs{\sigma}^{\Phi}(\mathbf{z}_i) - \mathbf{F}(\mathbf{z}_i) }_2^2,\label{eqn: pde loss}\\
        MSE_{bdry} &=\frac{1}{M_D}\sum_{i=1}^{M_D} \norm{\Phi(\hat{\mathbf{z}}_i) - \mathbf{g}(\hat{\mathbf{z}}_i)}_2^2+\frac{1}{M_N}\sum_{i=1}^{M_N} \norm{(\bs{\sigma}^\Phi \cdot\mathbf{n})(\bar{\mathbf{z}}_i) - \mathbf{h}(\bar{\mathbf{z}}_i)}_2^2,\\
        MSE_{init} &=\frac{1}{L} \sum_{i=1}^L \norm{\Phi(\tilde{\mathbf{z}_i}) - \mathbf{u}_0(\tilde{\mathbf{z}}_i)}_2^2 + \norm{\Phi_t(\tilde{\mathbf{z}_i}) - \mathbf{v}_0(\tilde{\mathbf{z}}_i)}_2^2,\\
        MSE &= MSE_{PDE} + MSE_{bdry} + MSE_{init},\label{eqn: second-order total MSE}
    \end{align}\label{eqn:MSE parts}%
\end{subequations}
for which the goal is to minimize \eqref{eqn: second-order total MSE} over the trainable network parameters of $\mathcal{N}$. 
Equations~\eqref{eqn:MSE parts} represents a general approximation for solving a second-order plane strain IBVP incorporating both hard and soft boundary enforcement. 

Our numerical tests in Section~\ref{sec: numerical tests} look at variations in network performance over different choices of Dirichlet boundary $\Gamma_D$ and Neumann boundary $\Gamma_N$ as well different choices for hard and soft boundaries $\Gamma, \overline{\Gamma}$ ranging from $\Gamma=\partial \Omega$ to $\overline{\Gamma}=\partial \Omega$. 
Furthermore, because hard-enforcement of traction conditions requires a network gradient in the trial function, see \eqref{eqn: 2nd order trial function} and \eqref{eqn: general normalizers}, the PDE loss function \eqref{eqn: pde loss} will require a third-order network derivative before backpropagation is even called to compute a network update. 
The nested derivatives that arise from the second-order problem lead us to also consider a first-order, velocity-stress formulation of the problem that will let us reduce both the order of the PDE and the order of the traction conditions.

\subsection{First-Order Form}
We may reformulate \eqref{eqn: 2nd-order governing eq} into first-order form by seeking a velocity-stress formulation $(\mathbf{v}, \bs{\tau}):\Omega\times[0,t] \to \mathbb{R}^5$. This is done by defining $\mathbf{v}=\mathbf{u}_t$ and $\bs{\tau}=\bs{\sigma}^\mathbf{u}$ so that \eqref{eqn: 2nd-order governing eq} becomes 
\begin{subequations}
    \begin{align}
        \rho \mathbf{v}_{t} - \nabla \cdot \bs{\tau} &= \mathbf{F},\\
        \bs{\tau}_t - \bs{\sigma}^\mathbf{v} &= \mathbf{S},\label{eqn: rate form Hooke}
    \end{align}\label{eqn: first-order governing eq}%
\end{subequations}
where \eqref{eqn: rate form Hooke} is the constitutive relation between strain rate and stress rate obtained by differentiating Hooke's law~\eqref{eqn: order2 hooke} with respect to time. We include $\mathbf{F,S}$ as numerical source terms so that we can verify accuracy using a manufactured solution which need not satisfy \eqref{eqn: first-order governing eq}(see Section~\ref{sec: first order numerics}). Boundary conditions are now specified on velocities and stresses to be
\begin{subequations}
    \begin{alignat}{3}
        \mathbf{v} &= {\mathbf{g}}(\mathbf{x}, t) &&\quad x\in \Gamma_D &&\quad t>0,\\
        \bs\tau\cdot \mathbf{n} &= {\mathbf{h}(\mathbf{x}, t)} &&\quad x\in \Gamma_N &&\quad t>0,\\
        \mathbf{v} &= {\mathbf{v}}_0(\mathbf{x}) &&\quad x\in \Omega &&\quad t=0,\\
        \bs{\tau} &= {\bs{\tau}}_0(\mathbf{x}) &&\quad x\in \Omega &&\quad t=0,
    \end{alignat}\label{eqn: first-order bc}%
\end{subequations}
which yields a well-posed IBVP for solving the plane strain equations. 
Again, denote the portions of the boundary where hard and soft enforcement are imposed, by $\Gamma$, and $\overline{\Gamma}$, respectively. 
Because we are approximating both velocities and stresses we must define a trial function for each solution space where relevant boundary conditions are enforced. 
For neural network $(\mathcal{V, T})$, let there be $K_1$ segments in $\Gamma\cap\Gamma_D$ and $K_2$ segments in $\Gamma\cap\Gamma_N$ each associated with approximate distance functions $\{\phi_{i}\}_{i=1}^{K_1}, \{\psi_{i}\}_{i=1}^{K_2}$, respectively. 
Using the same structure as in \eqref{eqn: 2nd order trial function} we define the trial function approximation $(\Phi, \Psi)\approx (\mathbf{v}, \bs{\tau})$ by
\begin{subequations}
    \begin{align}
        \Phi &= \sum_{i=1}^{K_1}f_iw_i + \mathcal{V} \prod_{i=1}^{K_1}\phi_i^{\eta_i},\label{eqn: velocity trial function}\\
        \Psi &= \sum_{i=1}^{K_2}q_ir_i + \mathcal{T} \prod_{i=1}^{K_2}\psi_i^{\xi_i},\label{eqn: stress trial function}
    \end{align}\label{eqn: first-order trial functions}%
\end{subequations}
where inverse distance weights $w_i, r_i$ are computed from $\{\phi_{i}\}_{i=1}^{K_1}, \{\psi_{i}\}_{i=1}^{K_2}$, respectively and exponents $\eta_i, \xi_i$ serve as controls on the interpolant's smoothness at boundary $\Gamma^\mathcal{V}, \Gamma^\mathcal{T}$, respectively (See Section~\ref{sec: hard-enforcement} for details). 
Since we impose conditions on the solution and not any of the solution derivatives, the normalizers are given simply as
\begin{align}
    f_i &= 
    \begin{cases}
        \mathbf{g} &\text{ if } \Gamma_i\subseteq \Gamma_D\\
        \mathbf{v}_0 &\text{ if } \Gamma_i\subseteq \Omega\times \{0\}
    \end{cases},\\
    q_i &= 
    \begin{cases}
        \mathbf{H} &\text{ if } \Gamma_i\subseteq \Gamma_N\\
        \bs{\tau}_0 &\text{ if } \Gamma_i\subseteq \Omega\times \{0\}
    \end{cases}
\end{align}
where, like before, we derive traction-compliant stress elements with respect to the network output. However, in the first-order problem stress is a solved quantity so we can enforce traction conditions as Dirichlet conditions on the stress variable. 
Thus, hard enforcement on tractions is achieved by the following:
\begin{subequations}
    \begin{align}
        s_{xy} &= (h_y - \mathcal{S}_{yy}n_y) / n_x,\\
        s_{xx} &= (h_x - s_{xy}n_y) / n_x, \\
        \mathbf{H} &= 
        \begin{bmatrix}
            s_{xx} & s_{xy}\\
            s_{xy} & \mathcal{S}_{yy}
        \end{bmatrix}.
    \end{align}
\end{subequations}
We may consider the same set of points as was used in the second-order loss formulation where we take interior data $\{\mathbf{z}_i\}_{i=1}^N\subset \Omega \times (0,t]$, Dirichlet boundary points $\{\hat{\mathbf{z}}_i\}_{i=1}^{M_D}\subset (\overline{\Gamma}\cap\Gamma_D)\times(0,t]$, traction boundary points $\{\bar{\mathbf{z}}_i\}_{i=1}^{M_N}\subset (\overline{\Gamma}\cap\Gamma_N)\times(0,t]$, and temporal boundary points $\{\tilde{\mathbf{z}}_i\}_{i=1}^{L}\subset \Omega \times \{0\}$.
Then the component mean squared error (MSE) loss functions corresponding to governing equations~\eqref{eqn: first-order governing eq} and boundary conditions \eqref{eqn: first-order bc} for approximate trial solution \eqref{eqn: first-order trial functions} are
\begin{subequations}
    \begin{align}
        MSE_{PDE} &= \frac{1}{N}\sum_{i=1}^N\norm{\rho\Phi_t(\mathbf{z}_i) -\nabla \cdot \Psi(\mathbf{z}_i) - \mathbf{F}(\mathbf{z}_i) }_2^2 + \norm{\Psi_t(\mathbf{z}_i) - \bs{\sigma}^\Phi(\mathbf{z}_i)-\mathbf{S}(\mathbf{z}_i)}_2^2,\label{eqn: first-order pde loss}\\
        MSE_{bdry} &=\frac{1}{M_D}\sum_{i=1}^{M_D} \norm{\Phi(\hat{\mathbf{z}}_i) - \mathbf{g}(\hat{\mathbf{z}}_i)}_2^2 + \frac{1}{M_N}\sum_{i=1}^{M_N} \norm{(\Psi \cdot\mathbf{n})(\bar{\mathbf{z}}_i) - \mathbf{h}(\bar{\mathbf{z}}_i)}_2^2,\\
        MSE_{init} &=\frac{1}{L} \sum_{i=1}^L \norm{\Phi(\tilde{\mathbf{z}_i}) - \mathbf{v}_0(\tilde{\mathbf{z}}_i)}_2^2+\norm{\Psi(\tilde{\mathbf{z}_i}) - \bs{\tau}_0(\tilde{\mathbf{z}}_i)}_2^2,\\
        MSE &= MSE_{PDE} + MSE_{bdry} + MSE_{init}.\label{eqn: first-order total MSE}
    \end{align}\label{eqn:first-order MSE parts}%
\end{subequations}
A solution to the first-order plane strain IBVP is found by minimizing \eqref{eqn: first-order total MSE} over the set of trainable network parameters that constitute the network $(\mathcal{V}, \mathcal{T})$. 
Compared to the second-order formulation, this form of the problem includes an additional constraint on the interior domain (Hooke's law) and an expanded solution space (velocity-stress) but also requires only one layer of automatic differentiation preceding backpropagation. 
Given both second and first-order forms of the problem we conduct numerical tests to observe the effects that the choice of boundary configuration has on the network's performance focusing on the achieved relative $L^2$-error of the network along with compilation and per-step training times.

\section{Numerical Tests}\label{sec: numerical tests}
\begin{figure}
    \centering
    \includegraphics[scale=1.0]{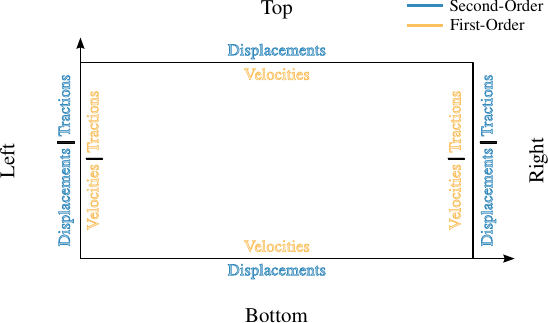}
    \caption{A domain schematic showing the types of boundary conditions considered at each boundary for first and second-order problems. Conditions at Left boundary are always chosen to be the same type as the conditions at the right boundary (e.g., Left and right both have displacement conditions or both have traction conditions when considering the second-order problem).}
    \label{fig: domain schematic}
\end{figure}

Each problem configuration in this section is solved using a fully-connected feed-forward neural network with three hidden layers, 128 neurons per layer, and swish activation functions. 
Network parameters are initialized using orthogonal initialization with a scaling factor of 0.9. 
Training is done with $1000$ uniformly random internal collocation points trained over $8000$ epochs with the L-BFGS optimizer. 
For soft-enforcement along a boundary, the first 125 internal collocation points are copied and projected on to the boundary for use in training. Table~\ref{tab: params} lists these specifications which are used for each numerical test.
Network initialization and collocation sampling are both seeded to ensure that each problem configuration is solved using the exact same network and collocation points. 
\begin{table}
    \centering
    \renewcommand{\arraystretch}{1.2}
    \scriptsize
    \begin{tabularx}{0.5\textwidth}{X X}
        Network Specifications & \\
        \toprule
        Architecture & FFNN\\
        Layers & 3\\
        Neurons & 128/layer\\
        Initialization & Orthogonal\\
        Orthogonal Scaling& 0.9\\
        Activation & Swish\\[1ex]
        Training Parameters& \\
        \toprule
        Algorithm & L-BFGS\\
        Internal points & 1000\\
        boundary points & 125 (projected)\\
        Epochs & 8000
    \end{tabularx}
    \caption{List of neural network parameters used across all numerical tests in this work.}
    \label{tab: params}%
\end{table}
We primarily explore the impact that varying soft and hard enforcement has on the network. 
To this end, we begin each simulation with only hard-enforced boundaries and progressively convert each boundary to soft-enforcement. 
This progression goes from all hard to all soft enforcement over 6 configurations, and progression is always done in the same way. 
Starting with all hard boundaries, we add soft boundaries in the following order: Left, Bottom, Right, Top, and the final soft boundary (time) is associated with all boundaries having soft-enforcement. 
Figures and tables will label each configuration by listing the initial letters of the soft boundaries (e.g., LBR denotes the three soft boundaries: Left, Bottom, and Right). Figure~\ref{fig: domain schematic} illustrates the various BC that we consider in this work.

\subsection{Second-Order Form}\label{sec: 2nd order numerics}%
We verify accuracy of our tests via the method of manufactured solutions \citep{Roache} using an established exact solution $u_e$ defined by 
\begin{align}
    u^e(x,y,t) = 
    \begin{bmatrix}
        \sin{(2\pi(x-t))}\\
        \sin{(2\pi(y-t))}
    \end{bmatrix},\label{eqn: manufactured sol}
\end{align}
which is used to derive source terms for governing equations~\eqref{eqn: 2nd-order governing eq} and boundary conditions~\eqref{eqn: 2nd-order bc}, namely 
\begin{subequations}
    \begin{alignat}{3}
        \mathbf{F} &= \rho u^e_{tt} - \nabla \cdot \bs{\sigma}^{u^e} &&\quad (x, y)\in \Omega, && t>0,\\
        \mathbf{g}(x, y, t) &= u^e(x, y, t), &&\quad (x, y)\in \Gamma_D, && t>0,\\
        \mathbf{h}(x, y, t) &= \bs{\sigma}^{u^e}(x, y, t) \cdot \mathbf{n}(x, y) , &&\quad (x, y)\in \Gamma_N, && t>0,\\
        \mathbf{u}_0(x, y) &= u^e(x, y, 0), &&\quad (x, y)\in \Omega, && t=0,\\
        \mathbf{v}_0(x,y) &= u_t^e(x, y, 0) , &&\quad (x, y)\in \Omega, && t=0.
    \end{alignat}
\end{subequations}
Accuracy is measured with respect to the $L^2-$norm over the entire spacetime domain $\hat{\Omega}=\Omega\times[0,1]$. Because the solution is a vector-valued function we employ the $L^2-$norm on a function space of vectors
\begin{align}
    \norm{f}_{2} = \left [ \int_{\hat{\Omega}} \norm{f(\mathbf{x})}^2 \, d\mathbf{x} \right]^{1/2}
\end{align}
where numerical integration is carried out using a composite Simpson's rule. We then define the relative $L^2-$error for the trial function \eqref{eqn: 2nd order trial function} by 
\begin{align}
    \norm{\Phi-u^e}_{rel} = \frac{\norm{\Phi-u^e}_{2}}{\norm{u^e}_{2}}.\label{eqn: order 2 relative L2 err}
\end{align}

\subsubsection{All Displacements}\label{sec: all disp}
\begin{figure}
    \centering
    \includegraphics[scale=0.75]{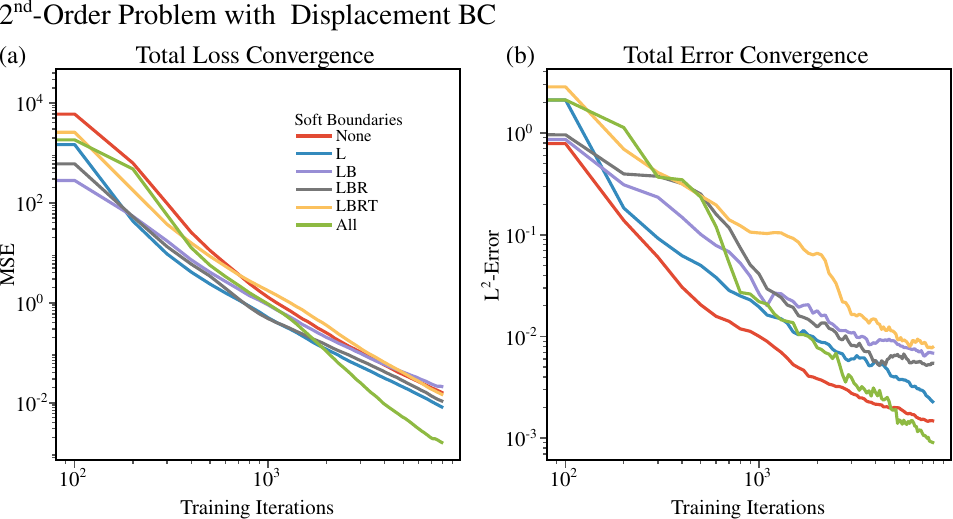}
    \caption{a) Total loss (MSE) and b) relative L2-error during network training when solving the Second-order plane strain problem with displacement conditions imposed at each spatial boundary. }
    \label{fig: order2_dirichlet}
\end{figure}

First we consider the scenario where each of the spatial boundaries are supplied with conditions on displacements, that is, the displacement boundary $\Gamma_D = \Gamma_L\cup\Gamma_B\cup\Gamma_R\cup\Gamma_T$ and the traction boundary $\Gamma_N=\emptyset$.
Figure~\ref{fig: order2_dirichlet} shows the convergence of the total loss~\eqref{eqn: second-order total MSE} and relative $L^2$-errors \eqref{eqn: order 2 relative L2 err} across varying configurations of hard and soft boundaries. 
This first case demonstrates that relatively smooth convergence in the loss function may be associated with noisy gradient updates suggesting that a decrease in total loss at a given training step need not correspond with a decrease in relative $L^2$-error. 
Moreover, we see in Figure~\ref{fig: order2_dirichlet} that when every boundary uses hard enforcement, the relative $L^2$-error shows that gradient updates tend to be less noisy than configurations that use soft enforcement. 
When solving the second-order problem with displacement conditions imposed on each spatial boundary we see similar relative accuracies achieved between the all-hard and all-soft configurations while configurations that mix both enforcement methods tended towards less accurate approximations. While the case of all-soft enforcement finishes training with the lowest relative error, the case of all-hard enforcement shows superior relative accuracy over fewer training iterations.

\subsubsection{Displacements and Tractions}
\begin{figure}
    \centering
    \includegraphics[width=0.75\linewidth]{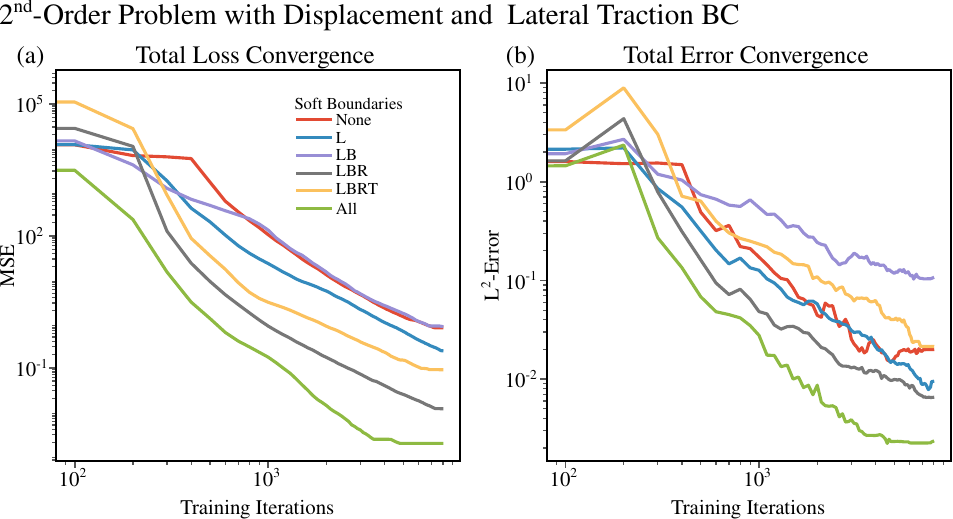}
    \caption{a) Total loss (MSE) and b) relative $L^2$-error during network training when solving the second-order plane strain problem with displacement conditions imposed at top and bottom boundaries and traction conditions imposed at left and right boundaries.}
    \label{fig: order2_tractions}
\end{figure}

Now, we repeat the same simulation as in Section~\ref{sec: all disp} but this time we specify traction conditions along the lateral boundaries so that displacement boundary $\Gamma_D = \Gamma_B\cup\Gamma_T$ while the traction boundary $\Gamma_N = \Gamma_L \cup \Gamma_R$. 
We report convergence data for total loss and relative $L^2$-error during network training in Figure~\ref{fig: order2_tractions}. 
Observe that early training iterations show little progress and even result in increasing $L^2$-error in configurations with soft enforcement. Once training picks up we see steady, albeit noisy, reduction in $L^2$-error. 
Comparing loss and $L^2$-error convergence in Figure~\ref{fig: order2_tractions}, we see that pure hard enforcement is outperformed by pure soft enforcement when looking at relative $L^2$-error but when comparing their final MSE we can see that the final loss value for the pure hard enforcement configuration is nearly three orders of magnitude larger than the final loss for pure soft enforcement. 
The relationship between magnitude of loss and relative $L^2$-error may be significant when training on larger data with more powerful hardware where losses may become small and training plateaus.

\subsection{First-Order Form}\label{sec: first order numerics}%
We use the manufactured solution in \eqref{eqn: manufactured sol} to define manufactured velocity $v^e$ and stress $\tau^e$ as
\begin{subequations}
    \begin{align}
        v^e(x, y, t) &= u_t^e(x, y, t)\\
        \tau^e(x, y, t) &= \bs{\sigma}^{u^e}(x, y, t)
    \end{align}
\end{subequations}
which may not satisfy the governing equations and constitutive law \eqref{eqn: first-order governing eq} so we compute appropriate source terms for each and specify boundary data accordingly. The resulting source terms are then given to be
\begin{subequations}
    \begin{alignat}{3}
        \mathbf{F}(x, y, t) & = \rho v^e_t(x, y, t) - \tau^e(x, y, t) &&\quad (x,y)\in \Omega, &&\quad t>0, \\
        \mathbf{S}(x, y, t) &= \tau_t^e(x, y, t) - \bs{\sigma}^{v^e}(x, y, t) &&\quad (x,y)\in \Omega, &&\quad t>0, \\
        \mathbf{g}(x, y, t) &= v^e(x, y, t) &&\quad (x,y)\in \Gamma_D, &&\quad t>0, \\
        \mathbf{h}(x, y, t) &= \tau^e(x, y, t) \cdot \mathbf{n}(x, y) &&\quad (x,y)\in \Gamma_N, &&\quad t>0, \\
        \mathbf{v}_0(x, y) &= v^e(x, y, 0) &&\quad (x,y)\in \Omega, &&\quad t=0, \\
        \bs{\tau}_0(x, y) &= \tau^e(x, y, 0) &&\quad (x,y)\in \Omega, &&\quad t=0. 
    \end{alignat}
\end{subequations}
Note that while we named boundaries $\Gamma_D, \Gamma_N$ to represent Dirichlet and Neumann boundaries, respectively, we must point out that in the first-order problem, all of the boundaries are Dirichlet boundaries because we solve for both velocity and stress. 
Therefore, we will use $\Gamma_D$ to refer to boundaries where we specify velocity conditions, and $\Gamma_N$ will denote the portion of the boundary where traction conditions are imposed on the stress variable. 

In first-order form, solutions take the form of velocities and stresses so we define the $L^2$-norm on a function space of direct products by
\begin{align}
    \norm{(w, \bs{s})}_{2} = \left [\int_{\hat{\Omega}} \norm{w(\mathbf{x})}^2 + \norm{\bs{s}(\mathbf{x})}^2_F \, d\mathbf{x} \right ]^{1/2}
\end{align}
where $\norm{\cdot}_F$ is the Frobenius norm and, again, numerical integration is carried out with a composite Simpson's rule. Relative $L^2-$error for the trial function~\eqref{eqn: first-order trial functions} is then given by
\begin{align}
    \norm{(\Phi-v^e, \Psi-\tau^e)}_{rel} = \frac{\norm{(\Phi-v^e, \Psi-\tau^e)}_{2}}{\norm{(v^e, \tau^e)}_{2}}.\label{eqn: order 1 relative L2 err}
\end{align}

\subsubsection{All Velocities}
\begin{figure}
    \centering
    \includegraphics[width=0.75\linewidth]{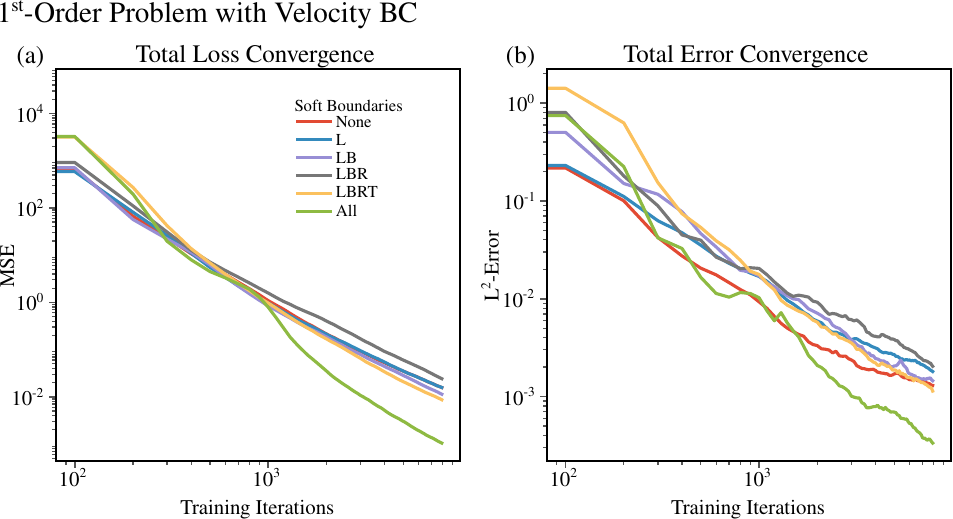}
    \caption{a) Total loss (MSE) and b) relative $L^2$-error during network training when solving the first-order plane strain problem with velocity conditions imposed at each spatial boundary.}
    \label{fig: order1_dirichlet}
\end{figure}

Continuing in a fashion similar to the second-order numerical tests, we will first consider the case where velocity is specified at every spatial boundary so that $\Gamma_D = \Gamma_L\cup\Gamma_B\cup\Gamma_R\cup\Gamma_T$ while the traction boundary $\Gamma_N = \emptyset$. 
Figure~\ref{fig: order1_dirichlet} shows convergence data for a network approximation to the first-order problem with velocity conditions at each spatial boundary measured across different configurations of hard and soft boundaries. 
This time we see comparable performance across each configuration with pure soft enforcement achieving the best relative $L^2$-error.

\subsubsection{Velocities and Tractions}
\begin{figure}
    \centering
    \includegraphics[width=0.75\linewidth]{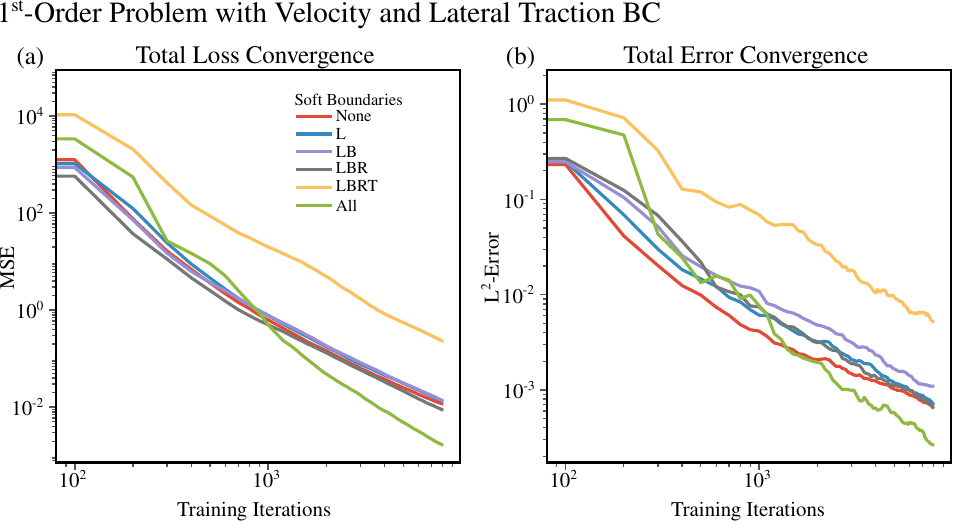}
    \caption{a) Total loss (MSE) and b) relative L2-error during network training when solving the first-order plane strain problem with velocity conditions imposed at top and bottom boundaries and traction conditions imposed on the stress variable at the left and right boundaries.}
    \label{fig: order1_tractions}
\end{figure}

Finally, we consider the case where velocities are specified at the top and bottom boundaries $\Gamma_D=\Gamma_B\cup\Gamma_T$ and tractions specified at right and left boundaries $\Gamma_N = \Gamma_L \cup \Gamma_R$. 
Figure~\ref{fig: order1_tractions} shows convergence data for a network approximation to the first-order problem with velocity conditions (at the top and bottom boundaries) and traction conditions (at left and right boundaries)  measured across different configurations of hard and soft boundaries. 
To account for performance variation across different data sets, we perform each numerical test over ten different sets of $1000$ random collocation points and present average relative $L^2$-errors in Table~\ref{tab:avg rel L2}. 
\begin{table}[t]
    \centering
    \renewcommand{\arraystretch}{1.2}
    \scriptsize
    \begin{tabularx}{0.8\textwidth}{l *{4}{S[table-format = 2.3e2]}}
        \multicolumn{5}{l}{\normalsize  Relative L2-Error Across Boundary Configurations} \\ [1ex]
        \toprule \toprule 
        &\multicolumn{2}{c}{Second-Order Form}& \multicolumn{2}{c}{First-Order Form}   \\
        \cmidrule(lr){2-3}\cmidrule(lr){4-5}
        {soft bdrys}&    {Displacement Only}  & {Displacement $+$ Traction}      &{Velocity Only}  & {Velocity $+$ Traction} \\
        \midrule
        All	&	5.676e-03	&	1.986e-03	&	5.238e-04	&	3.690e-04\\
        	&	[3.588e-03] 	&	[1.038e-03]	&	[3.655e-04] 	&	[1.003e-04]\\[1ex]
        LBRT	&	6.428e-02	&	3.595e-02	&	1.192e-03	&	1.604e-03\\
        	&	[2.343e-02] 	&	[1.555e-02]	&	[3.741e-04] 	&	[6.294e-04]\\[1ex]
        LBR	&	3.255e-02	&	3.267e-02	&	1.639e-03	&	1.256e-03\\
        	&	[1.169e-02] 	&	[3.480e-02] 	&	[4.672e-04] 	&	[4.744e-04]\\ [1ex]
        LB	&	7.475e-02	&	1.805e-01	&	2.091e-03	&	1.497e-03\\
        	&	[7.164e-02] 	&	[8.872e-02]	&	[5.464e-04] 	&	[4.078e-04] \\[1ex]
        L	&	1.891e-02	&	4.020e-02	&	1.788e-03	&	1.448e-03\\
        	&	[9.726e-03]	&	[4.206e-02] 	&	[5.082e-04] 	&	[2.243e-04]\\ [1ex]
        None	&	1.364e-03	&	5.117e-02	&	1.505e-03	&	1.868e-03\\
        	&	[5.596e-04]	&	[3.947e-02] 	&	[6.388e-04]	&	[1.202e-03] \\
        \bottomrule
    \end{tabularx}
    \caption{Relative $L^2$-norm associated with network solutions for first and second-order plane strain problems each solved for two different sets of boundary conditions. Variation in performance is recorded over different configurations of hard and soft boundary enforcement. Variation in performance is accounted for by averaging final training accuracy across 10 different sets of collocation points. Bracketed values placed below each reported mean give the associated standard deviation.}
    \label{tab:avg rel L2}%
\end{table}
As indicated in the table, the PINN achieves better accuracy when solving the first-order problem. Indeed, for all of our tests we found that using soft-enforcement at every boundary tended to yield a more accurate solution over mixed-enforcement or pure hard-enforcement. 
However, considering the increased accuracy observed when solving the first-order problem we present timing metrics in Table~\ref{tab: timing metrics} to get a more complete picture. 
\begin{table}
    \centering
    \renewcommand{\arraystretch}{1.2}
    \scriptsize
    \begin{tabularx}{0.8\textwidth}{l *{6}{X}}
        \multicolumn{5}{l}{\normalsize  Timing Metrics for Solving First-Order Problem } \\ [1ex]
        \toprule \toprule 
        &\multicolumn{3}{c}{Velocity only}& \multicolumn{3}{c}{Velocity and Tractions}   \\
        \cmidrule(lr){2-4}\cmidrule(lr){5-7}
        soft boundaries& Total & Compilation & Update & Total & Compilation & Update(\si{\milli\second})\\
         & (\si{\second})& (\si{\second}) & (\si{\milli\second}) & (\si{\second}) & (\si{\second}) & (\si{\milli\second})\\
        \midrule
        All	&	91.7	&	5.14(5.6\%)	&	10.82	&	95.09	&	5.73(6.0\%)	&	11.17\\
        	&	[18.41]	&	[0.62]	&	[2.32]	&	[14.00]	&	[0.82]	&	[1.70]\\[1ex]
        LBRT&	63.4	&	6.15(9.7\%)	&	7.16	&	50.3	&	6.88(13.7\%)	&	5.43\\
        	&	[20.88]	&	[0.91]	&	[2.55]	&	[14.36]	&	[1.06]	&	[1.82]\\[1ex]
        LBR	&	42.7	&	7.67(18.0\%)	&	4.38	&	43.52	&	8.13(18.7\%)	&	4.42\\
        	&	[8.24]	&	[1.05]	&	[0.92]	&	[6.32]	&	[1.15]	&	[0.81]\\[1ex]
        LB	&	37.7	&	8.82(23.4\%)	&	3.61	&	42.66	&	10.63(24.9\%)	&	4.00\\
        	&	[1.64]	&	[1.14]	&	[0.15]	&	[6.00]	&	[1.60]	&	[0.59]\\[1ex]
        L	&	40.8	&	10.32(25.3\%)	&	3.81	&	42.88	&	11.17(26.0\%)	&	3.96\\
        	&	[2.71]	&	[1.18]	&	[0.32]	&	[6.01]	&	[1.62]	&	[0.75]\\[1ex]
        None&	50.32	&	10.85(21.6\%)	&	4.94	&	42.33	&	12.84(30.3\%)	&	3.69\\
        	&	[7.81]	&	[1.17]	&	[0.98]	&	[2.39]	&	[1.74]	&	[0.25]\\[1ex]
        \bottomrule
    \end{tabularx}
    \caption{Total run time, compile time and per iteration network update time for the PINN solution for the first-order plane strain problem considering the case where only velocity conditions are imposed on spatial boundaries and the case where velocity conditions are used on top and bottom boundaries with traction conditions on left and right boundaries. Compile time is given in seconds followed by a percent showing how much of total run time was due to compilation. Results are averaged over 10 different sets of 1000 collocation points. Bracketed values below each reported mean give the associated standard deviation.}
    \label{tab: timing metrics}%
\end{table}
There we see that compilation time decreases as the number of soft boundaries increases and per-iteration network update time increases as the number of soft boundaries increases. 
However, over many training epochs, the per-iteration update time accumulates and can lead to longer total run times. 
Implementing hard-enforcement at every boundary yields run times that are nearly half that of implementing soft-enforcement at every boundary.

Our findings highlight an important trade-off between the choice of hard and soft boundary enforcement. 
Hard-enforcement may yield longer compilation times (due to compiling a more complex trial function), but reduces the time for each training step.  
Soft-enforcement can reduce compilation time, but may increase the execution time of each training step update. 
Thus, when building forward models it is important to consider weather accuracy or run time should be prioritized.

\section{Summary and future work}
\label{sec: summary and future work}
We have presented a generalized approach for both hard and soft-enforcement of boundary conditions in a physics-informed neural network solution to the elastodynamic plane strain problem. 
This approach is carried out for both first and second-order formulations of the wave equation. We demonstrate the proper normalizations for boundary data for both Dirichlet and Neumann boundary types and we provide a computational framework that extends to other problems in classical continuum mechanics . 
In the second-order problem we demonstrate how to enforce conditions on displacement along with conditions on the Jacobian of displacement that result in traction compliance at a boundary. 
In the first-order problem all of the boundary conditions become Dirichlet type and the enforcement on velocity and stress variables yield traction-compliance at the boundaries. 
We then solve each problem formulation considering two different lateral boundary types across six configurations of soft and hard boundary enforcement and test performance using the method of manufactured solutions. 

We find that the PINN performs better in the first-order formulation, likely due to the reduced differential order of the problem for both governing equations and traction boundaries. Furthermore, the second-order problem with traction conditions is where the PINN performs worse, likely due to the higher-order governing equation and nested derivative that arises in the traction normalizer. 
We observe that mixing soft and hard enforcement tends to yield errors on the same order as only using hard-enforcement. 
This suggests that the relative $L^2$-error is dominated by distance approximation on the interior of the domain and warrants further investigation. 
When looking at execution metrics for solving the first-order problem we show that the final accuracy of the trained network may be related to total run time via the choice of boundary enforcement method. 
Longer run are times associated with soft boundary enforcements and could achieve higher relative accuracy compared to the shorter run times associated with hard boundary enforcements. 

We plan to extend this work to feature additional network architecture for inferring frictional properties that inform seismic stability along earthquake faults. Our aim is to construct a numerical analog for the laboratory experiments for deriving fault friction relationships \citep{ Cebry2022, Mclaskey2017, McLaskey2019} and using implicit boundary representations to explore the effects of more complex fault geometries as in \citet{cebry2024heterogeneous}. 
The next steps will be a twofold exploration; verifying PINN inference on laboratory data, and predicting laboratory slip events. Such a study could also consider the difference between inferring friction directly from the data versus inferring parameters of a governing friction law. 
Further development to resolve longer time scale simulations may be necessary to implement this approach in community benchmark problems concerning dynamic rupture simulations \cite{Harris2009} and sequences of earthquakes and aseismic slip \cite{Erickson2020}.

\newpage
\appendix

\section{Notes on R-function Theory: Implicit boundary representation}\label{sec: R-function theory}
Normalized, implicit functions can be constructed with guaranteed differential properties using the theory of R-functions \citep{shapiro1991theory, shapiro1999implicit}. Here we cover some of the basics of R-function theory and show how it is used to construct implicit representations of semi-analytic sets. For the purposes of this work, we provide a brief introduction to R-functions which we then use to construct the requisite ADFs for interpolating data at domain boundaries. 

A real-valued function $F(\omega_1, \omega_2, \dots, \omega_q)$, where $\{\omega_i: \mathbb{R}^n\mapsto \mathbb{R}\}_{i=1}^q$, is an R-function if the sign of $F(\,\cdot\,)$ is completely determined by the sign of its arguments $\omega_i(\mathbf{x})$\citep{shapiro1991theory, rvachev2001transfinite}. Because $F$ only changes sign when its arguments change sign, it operates like a Boolean function where positive and negative outputs are associated with logical true and logical false, respectively. R-functions are then closed under composition (just like Boolean functions) and we can construct any $R$-function from a small number of non-unique elementary R-functions \citep{shapiro2007semi}. Like their Boolean counterparts, R-negation$(-\omega)$, R-disjunction$(\omega_1\lor \omega_2)$, and R-conjunction$(\omega_1 \land \omega_2)$ correspond to the set operations of complement, union, and intersection, respectively. However, any given logic function is a companion to infinitely many R-functions so that we may choose an appropriate system of R-functions for our application.

Consider a function $f\equiv \omega_1 \tilde{\land}\omega_2$ where $\omega_1, \omega_2$ are any real-valued functions. Then $f$ is positive if and only if both $\omega_1$ and $\omega_2$ are positive, that is,
\begin{align}
    (\omega_1 \tilde{\land}\omega_2)\geq 0 \iff (\omega_1\geq 0) \land (\omega_2\geq 0).
\end{align}
This demonstrates the main result of the theory of R-functions: A single implicit function can be used to represent any region defined by some logical predicate on a system of inequalities of the same type \citep{rvachev2001transfinite, shapiro1991theory}. This translation is done via simple syntactic replacement by swapping set operations for their corresponding R-function and set representatives for half-space primitive functions. 

If $\omega(\mathbf{x})=0$ defines some $n$-dimensional hypersurface $\Omega$ in $\mathbb{R}^n$ then the zero-curve of $\omega$ is also represented by $-\vert \omega\vert\geq 0$. For a trim region $T$ defined by $\tau\geq0$ such that $T \cap \Omega \neq \emptyset$, a \textit{trimmed} hypersurface is defined by $\Omega \cap T$. Then, by syntactic replacement, the trimmed hypersurface is given by $\left (-\vert\omega \vert \tilde{\land} \tau \right )\geq  0$. This technique produces a bounded, implicit curve from the (potentially) unbounded curve $\omega(\mathbf{x})=0$ and bounding region $\tau\geq0$. Then, to construct an ADF from from some boundary $\Gamma$, we need construct an implicit representation of $\Gamma$ and an appropriate trimming region.  


An implicit function $\omega(\mathbf{x})=0$ can be represented by the inequality $-\sqrt{\omega}\geq 0$. In general, the distance from the implicit curve $\omega(\mathbf{x})=0$ is not the same as the distance from a finite segment of the curve. Because of this, we must trim the implicit curve 

\section{Approximate distance functions for Bezier curves}\label{sec: A bezier}
\subsection{Implicit Bezier curve}
Given a set of control points $P\in\mathbb{R}^{2\times n}$ the Bezier curve of degree $n-1$ parametrized by $t\in [0,1]$ is given by
\begin{align}
    \gamma(t) = PM\mathbf{t}\label{eqn: bezier def}
\end{align}
where $\mathbf{t}=[t^{n-1}, \cdots,t, 1]^T$ and $M\in \mathbb{R}^{n\times n}$ is derived from the Bernstein basis to be
\begin{align}
    M_{ij} = \begin{cases}
        \binom{n}{i}\binom{n-i}{j}(-1)^{n+i-j}, &\quad 0\leq i \leq n, 0\leq j\leq n-i\\
        0, &\quad \text{otherwise}
    \end{cases}.
\end{align}
Equation~\eqref{eqn: bezier def} provides, explicitly, a polynomial representation for the $x,y-$coordinates in $\mathbb{R}^2$. It is useful to consider each component polynomial separately so we let $PM=\begin{bmatrix}\mathbf{a} & \mathbf{b}\end{bmatrix}^T$ where $\mathbf{a}$, $\mathbf{b}$ are the coefficients associated with polynomial coordinates $x$ and $y$, respectively. Then, the implicit function

\begin{align}
\Gamma(x, y) = \begin{bmatrix}
        a_{n-1} & a_{n-2} & \cdots & a_1 & a_0-x\\
        b_{n-1} & b_{n-2} & \cdots & b_1 & b_0-y
    \end{bmatrix}
    \begin{bmatrix}
        t^{n-1}\\
        \vdots\\
        t\\
        1
    \end{bmatrix}
    =\begin{bmatrix}
        0\\0
    \end{bmatrix}
\end{align}
defines an implicit Bezier curve. From elimination theory we can construct an expression, known as a resultant, out of the polynomial coefficients such that the resultant vanishes if and only if the set of polynomials have a common, nontrivial root\citep{sederberg1984implicit}. Here we will use the approach presented in \citet{chionh2002fast} for fast computation of the Bezout resultant.
By defining auxiliary coefficient matrices $L, R$ by
\begin{align}
    L_0 = 
    \begin{bmatrix}
        a_0-x & b_0 - y
    \end{bmatrix}, \quad 
    L_i = 
    \begin{bmatrix}
        a_i & b_i
    \end{bmatrix}, \quad
    R_i = 
    \begin{bmatrix}
        b_i & -a_i
    \end{bmatrix}^T, \quad i=1, 2, \dots, n
\end{align}
each row of the Bezout resultant matrix can be computed by
\begin{alignat}{2}
    B_{0j} &= L_0R_i,  &&\quad 1\leq j\leq n, \\
    B_{ij} &= B_{i-1, j+1} + L_iR_{j+1}, &&\quad i\leq j\leq n-2,\\
    B_{nj} &= L_{n-1}R_n, &&\quad 1\leq j\leq n,
\end{alignat}
where symmetry $B_{ij}=B_{ji}$ is used to only compute the upper triangle of the resultant matrix. Finally, the approximate distance from the Bezier curve $\gamma$ is given by the resultant determinant
\begin{align}
    \Gamma(x, y) = \lvert B \rvert.\label{eqn: implicit bez}
\end{align}
While \eqref{eqn: implicit bez} is an approximate distance we normalize $\Gamma$ so that $\lvert \lvert \nabla \Gamma \rvert \rvert=1$ in a neighborhood of $\Gamma =0$ for a higher-order approximation of distance\citep{upreti2014algebraic}. This is done with the first-order normalization 
\begin{align}
    \widetilde{\Gamma} = \frac{\Gamma}{\sqrt{\Gamma^2+ \left\lvert\lvert \nabla \Gamma \right \rvert\rvert^2}}\label{eqn: normalize implicit bezier}
\end{align}
given in \citet{shapiro2007semi}. Finally, we note that $\widetilde{\Gamma}$ is an unbounded, implicit curve in $\mathbb{R}^2$ which can impact its function as a distance approximation. To remedy this, we want to restrict $\widetilde{\Gamma}$ to the bounded segment defined in \eqref{eqn: bezier def}. This is done by trimming the implicit function with the Bezier control polygon.

\subsection{Bezier Trimming Region}
\begin{figure}
    \centering
    \includegraphics[width=\linewidth]{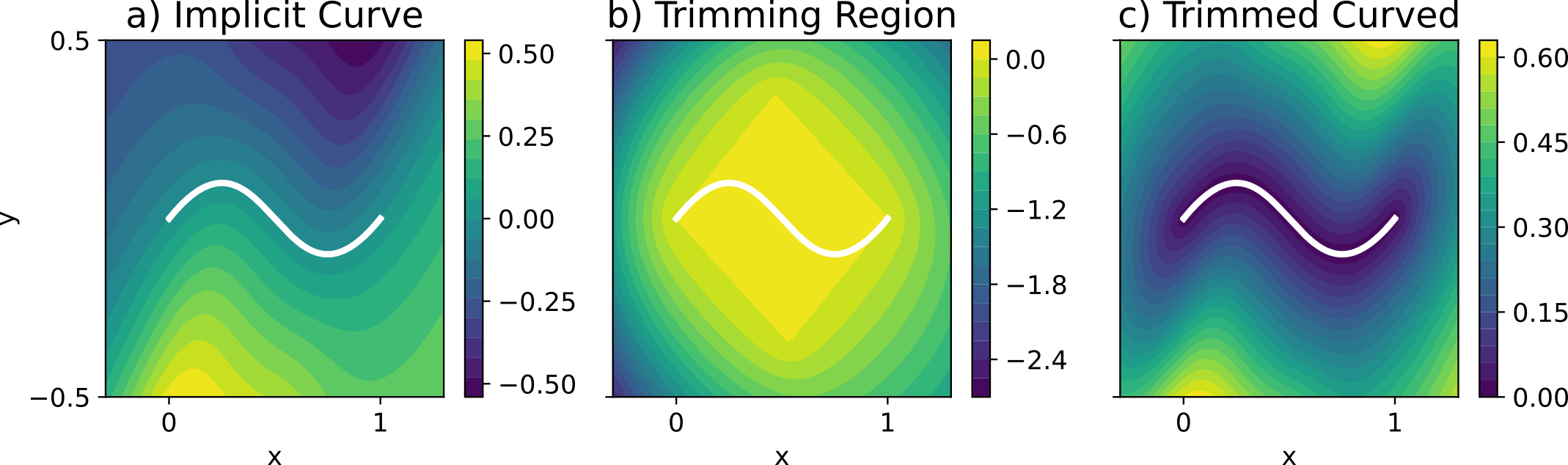}
    \caption{Illustration of the a) implicit Bezier curve, b) the trimming region formed from the convex hull of Bezier control points, c) the resulting trimmed Bezier curve.}
    \label{fig: bezier adf parts}
\end{figure}

\begin{figure}
    \centering
    \includegraphics[width=1.0\linewidth]{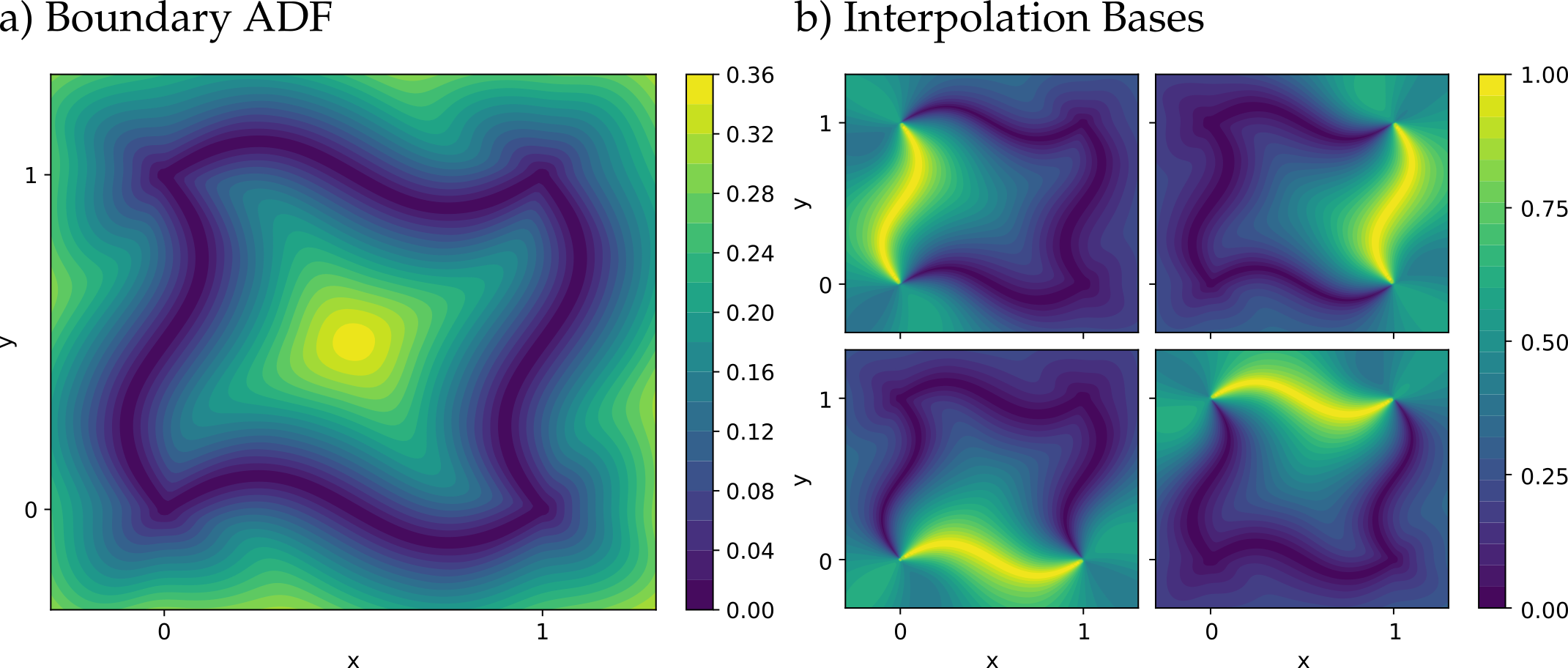}
    \caption{Illustration of a domain using four Bezier curves as boundaries and the resulting a) interior adf to the entire boundary and b) interpolation bases computed from inverse distance weighting.}
    \label{fig: bezier interpolating bases}
\end{figure}

The Bezier trimming region is determined by the convex hull of the Bezier control points. Here, we will use the implicit functions defined in \eqref{eqn: implicit line} to construct signed distances for each edge of the Bezier control polygon. Orientation is important here since the implicit line has two possible outcomes for any given pair of points. Because we are constructing a trimming function, the inequality $t\geq0$ dictates that each signed distance should be oriented such that points in the curve's convex hull are a nonnegative distance from the polygonal edge. Let $\{\gamma_i\}_{i=1}^n$ be the implicit line that contains edge $i$ of the convex hull of Bezier control points. Each pair of lines can be joined by $R$-conjunction $\phi_s$ defined by 
\begin{align}
    \phi_s(\phi_1, \phi_2) \coloneq  \phi_1 + \phi_2 - \sqrt[s]{\phi_1^s + \phi_2^s}
\end{align}
which is normalized up to order $(s-1)$\citep{sukumar2022exact}.  For $k=2, 3, \dots,n$ the $kth$ joining of edges is defined recursively by 
\begin{align}
    t_2 &= \phi_s(\gamma_1, \gamma_2)\\
    t_k &= \phi_s(t_{k-1}, \gamma_k) 
\end{align}
and the convex trimming region $t\geq 0$ defined by the convex hull of Bezier control points is given by 
\begin{align}
    t\coloneq t_n \label{eqn: bezier hull trim}
\end{align}
Then, the ADF for the trimmed curve segment is obtained by trimming \eqref{eqn: normalize implicit bezier} with \eqref{eqn: bezier hull trim} using \eqref{eqn: normalized trim}. Figure~\ref{fig: bezier adf parts} shows the implicit Bezier curve \eqref{eqn: normalize implicit bezier}, convex trimming region \eqref{eqn: bezier hull trim} and the resulting trimmed curve.
Figuere~\ref{fig: bezier interpolating bases} shows the resulting interior ADF and the interpolation bases formed from inverse distance weighting using four Bezier curves as boundaries.

\let\url\nolinkurl
\bibliographystyle{elsarticle-harv}
\bibliography{main}

\end{document}